\documentclass{emulateapj}
\usepackage{apjfonts}
\usepackage{amsmath}
\usepackage{natbib}
 
\newcommand{\bi}[1]{\ensuremath{\boldsymbol{#1}}} 
\slugcomment{accepted for publication in ApJ}
\shorttitle{Two-Component Outflow Powered by Magnetic Explosion on Compact Stars}
\shortauthors{Matsumoto et al.}
\begin{document}
\title{Special Relativistic Magnetohydrodynamic Simulation of Two-Component Outflow Powered by Magnetic Explosion on Compact Stars} 
\author{Jin Matsumoto\altaffilmark{1,2}, Youhei Masada\altaffilmark{3,4}, Eiji Asano\altaffilmark{1}, and Kazunari Shibata\altaffilmark{1}}
\altaffiltext{1}{Kwasan and Hida Observatories, Kyoto University, Kyoto, Japan; jin@kusastro.kyoto-u.ac.jp}
\altaffiltext{2}{Department of Astronomy, Kyoto University, Kyoto, Japan}
\altaffiltext{3}{Graduate School of System Informatics, Department of Computational Science, Kobe University, Kobe, Japan}
\altaffiltext{4}{Hinode Science Project, National Astronomical Observatory of Japan, Tokyo, Japan}
\begin{abstract}
The nonlinear dynamics of outflows driven by magnetic explosion on the surface 
of a compact star is investigated through special relativistic magnetohydrodynamic 
simulations. We adopt, as the initial equilibrium state, a spherical stellar object embedded 
in hydrostatic plasma which has a density $\rho(r) \propto r^{- \alpha}$ and is 
threaded by a dipole magnetic field. The injection of magnetic energy at the surface 
of compact star breaks the equilibrium and triggers a two-component outflow. 
At the early evolutionary stage, the magnetic pressure increases rapidly around 
the stellar surface, initiating a magnetically driven outflow. A strong forward shock 
driven outflow is then excited. The expansion velocity of the magnetically driven 
outflow is characterized by the Alfv\'en velocity on the stellar surface, and follows 
a simple scaling relation $v_{\rm mag} \propto {v_{\rm A}}^{1/2}$. When the initial 
density profile declines steeply with radius, the strong shock is accelerated self-similarly 
to relativistic velocity ahead of the magnetically driven component. We find that it 
evolves according to a self-similar relation $\Gamma_{\rm sh} 
\propto r_{\rm sh}$, where $\Gamma_{\rm sh}$ is the Lorentz factor of the plasma 
measured at the shock surface $r_{\rm sh}$. Purely hydrodynamic process would 
be responsible for the acceleration mechanism of the shock driven outflow. Our 
two-component outflow model, which is the natural outcome of the magnetic explosion, 
can provide a better understanding of the magnetic active phenomena on various magnetized 
compact stars.
\end{abstract}
\keywords{MHD, relativistic processes, stars: neutron - stars: winds, outflows, methods: numerical}
\section{Introduction}
Plasma outflow from gravitationally bounded stellar objects is a universal 
phenomenon in astrophysics which is known to occur on a range of different 
spatial and energetic scales. Gaseous material containing synthesized elements 
and energy are supplied from the stellar source into the surrounding environment 
by these outflows. It is important to study the driving mechanism of the plasma outflow 
not only for the dynamical evolution of the stellar object itself, 
but also for the chemical and dynamical evolution of interstellar medium of galaxies.  

The pioneering theory of stellar wind, which is the prototype of the astrophysical 
outflow, was developed by Parker (1958). It describes steadily accelerated plasma 
flow from a high-temperature corona into interstellar space. Weber \& Devis (1967) 
extended Parker's wind theory to include magnetic processes by modeling plasma 
outflows along open magnetic field line. In this model,  the accelerated stellar wind 
collides with the interstellar medium and finally forms the shock wave.

The neutron star is known to have a relativistic plasma outflow, called as pulsar wind.  
The magnetosphere of rotating magnetized neutron star is filled with charged  
particles (Goldreich \& Julian 1969; Sturrock 1971; Ruderman \& Sutherland 1975). 
These charged particles flow out steadily along open magnetic field lines through the 
light cylinder where the co-rotation velocity coincides with the light speed. The plasma 
is accelerated to relativistic velocity by converting the rotational energy of the pulsar itself.
The pulsar wind also interacts with the surrounding dense gas ejected from the progenitor 
stars and causes a termination shock (Kennel \& Coroniti 1984).

Non-steady outflows exist universally in astrophysical systems. These should be seen 
as a more general class of outflows than the steady stellar wind. Of these non-steady flows, 
astrophysical jets are the most studied. Astrophysical jets are collimated bipolar outflows 
from central objects. It is believed that magneto-centrifugal and magnetic pressure  
forces are the most promising power for driving the collimation flow (Blandford \& Payne 1982; 
Uchida \& Shibata 1985; Shibata \& Uchida1986). The magnetized jet from young stellar 
objects have a central role in prompting the collapse of molecular clouds and thus is  
essential for star formation (Machida et al. 2008). The super-massive black holes in active 
galactic nuclei also powers bipolar jets (Begelman et al. 1984; Bridle \& Perley 1984). 
The most energetic event in the universe is the gamma-ray burst, and which is believed 
to be energized by ultra-relativistic bipolar flows formed when a massive stellar core 
collapses (Woosley 1993; MacFadyen \& Woosley 1999).

The non-steady mass ejection event associated with solar flares is called "coronal
mass ejection (CME)" (Low 1996). The emergence of a twisted magnetic flux tube 
from below the solar surface, which supplies free energy to the system, triggers flares 
by magnetic reconnection (Ishii et al. 1998; Kurokawa et al. 2002;  Magara 2003) 
and drives the associated CMEs (e.g., Mikic \& Linker 1994; Shibata et al. 1995).

We can observe non-steady mass ejection phenomena analogous to the solar flare/CME 
activity in magnetar systems. The magnetar is a ultra-strongly magnetized neutron star 
with magnetic field of $10^{14}$--$10^{15}$ G (Duncan \& Thompson 
1992, Thompson \& Duncan 1995; Harding \& Lai 2006; Mereghetti 2008). The giant flare observed from 
the magnetar is surprisingly energetic ($10^{44}$--$10^{46}$ ergs) and is believed to 
be powered by releasing the enormous magnetic energy stored in the magnetar itself 
(Thompson  \& Duncan 1995; Lyutikov 2006; Masada et al. 2010). We note that the 
magnetar giant flare is also associated with an expanding ejecta (Cameron et al. 2005; 
Gaensler et al. 2005; Taylor et al. 2005).

While the energetic origin of the outflow is different in the various astrophysical 
systems, there is a characteristic common to all outflow phenomena. The magnetohydrodynamic 
(MHD) processes play a role in powering and regulating the outflow. It has been a central 
issue in the study of the astrophysical outflow to specify the MHD process which controls 
the outflow dynamics (Hayashi et al. 1996; Beskin \& Nokhrina 2006; Spruit 2010). Nevertheless, 
the physical property and the acceleration mechanism of the outflow powered by the magnetic 
energy released from the compact stars have not yet been studied sufficiently 
(e.g., Komissarov \& Lyubarsky 2004; Komissarov 2006; Takahashi et al. 2009).

This paper focuses on the outflow powered by a magnetic explosion on a compact 
star, such as neutron star. We investigate its nonlinear dynamics using 2.5-dimensional special 
relativistic MHD simulations. The purpose of this paper is to reveal the characteristics 
of the outflow which expands from the surface of the compact star into the surrounding 
medium. The relativistic effect on the physical properties of the outflow is a special interest 
of this work because such outflow should be easily accelerated to relativistic velocity.

This paper opens with descriptions of the numerical model in \S~2. The numerical results 
of our relativistic MHD simulation are presented in \S~3. The self-similar property of the 
outflow is studied in more detail using one-dimensional hydrodynamic simulation in \S~4. 
In \S~5, the characteristics and potential application of our outflow model are discussed.
Finally, we summarize our findings in \S~6.
\section{Numerical Model} 
\subsection{Governing Equations}
We solve the special relativistic magnetohydrodynamic (SRMHD) equations in an 
axisymmetric spherical polar coordinate system ($r$, $\theta$, $\phi$). The fluid velocity 
$\bold{v}$, electric field $\bold{E}$ and magnetic field $\bold{B}$ in the system have 
three components while their derivatives in the $\phi$-direction are assumed to be zero.
We adopt, as the equation of state for the surrounding medium, the ideal gas law with 
a ratio of the specific heats $\gamma = 4/3$. For simplicity all the dissipative and radiative 
effects are ignored. The basic equations are then
\begin{eqnarray}
\frac{\partial D}{\partial t}&+&\nabla \cdot (D \bold{v})=0 \;, \label{eq: mass conservation} \\ 
\frac{\partial \bold{R}}{\partial t}+\nabla \cdot \Biggl [ \biggl(P&+&\frac{B^2+E^2}{8\pi}\biggr)\bold{I} \nonumber \\
+\Gamma^2(e+P)\frac{\bold{v}\bold{v}}{c^2}&-&\frac{\bold{B}\bold{B}+\bold{E}\bold{E}}{4\pi}\Biggr ] =\rho \bold{g} \;, \label{eq:momentum conservation} \\
\frac{\partial \epsilon}{\partial t} &+& \bold{\nabla}\cdot \bold{R}c^2=\rho \bold{v} \cdot \bold{g} \;, \label{eq: energy conservation} \\
\frac{\partial\bold{B}}{\partial t}-\nabla \times (&\bold{v}& \times \bold{B})=0 \;, \label{eq: induction eq} \\
\bold{E}=&-&\frac{\bold{v} \times \bold{B}}{c} \:,
\end{eqnarray}
\noindent
where
\begin{eqnarray}
D&\equiv & \Gamma \rho \;, \\
\bold{R}\equiv\Gamma^2(e+ &P& ) \frac{\bold{v}}{c^2}+\frac{\bold{E}\times\bold{B}}{4\pi c} \;, \\
\epsilon\equiv\Gamma^2(e+P)- &P&+\frac{B^2+E^2}{8\pi}.
\end{eqnarray}
Here $\Gamma \equiv [1-(v/c)^2]^{-1/2}$ is the Lorentz factor, and $e \equiv \rho c^2 
+ P/(\gamma - 1) $ is the internal energy. 
In Newtonian gravity $\bold{g}$, we do not take account of the relativistic correction 
on the gravitational force because it should be negligible in the neutron star system. 
D, $\bold{R}$ and $\epsilon $ denote the density, the momentum density and the energy 
density measured in the laboratory frame respectively. The other symbols have their 
usual meanings.

We use the HLL scheme (Harten et al. 1983) to solve the SRMHD equations in the 
same manner as Leismann et al. (2005). The numerical flux across the cell interface is 
evaluated using the characteristic velocity of the fast magneto-sonic wave obtained from 
the physical variables in left and right adjacent cells of the interface. The primitive variables 
are calculated from the conservative variables following the method of Del Zanna et al. (2003). 
We use a MUSCL-type interpolation method to attain second order accuracy in space, 
and Constrained Transport method to guarantee a divergence free magnetic field 
(Evans \& Hawley 1988).
\subsection{Initial Setting of Numerical Model}
For our initial conditions, we consider a hydrostatic gas surrounding a central compact star. 
A dipole magnetic field anchored in the central star threads the external gaseous plasma, 
and is given by
\begin{eqnarray}
B_r &=& \frac{2 B_0 {R_{*}}^3 cos \theta}{r^3} \;, \label{eq: dipole Br} \\
B_{\theta} &=& \frac{B_0 {R_{*}}^3 sin \theta}{r^3} \;, \label{eq: dipole Btheta}\\
B_{\phi} &=& 0.
\end{eqnarray} 
where $B_0$ and $R_*$ are the field strength at the equatorial plane
on the stellar surface and the radius of the central star respectively.
We assume that the temperature of the surrounding medium is
inversely proportional to the distance from the centroid of the star
\begin{eqnarray}
T = T_0 \biggl ( \frac{r}{R_{*}} \biggr )^{-1} \;, \label{eq: initial T}
\end{eqnarray}
where $T_0$ is the reference temperature of the surrounding gas on the surface of the
central star. Then the spatial distribution of the gas density and pressure in hydrostatic 
equilibrium are
\begin{eqnarray}
\rho &=& \rho_{0}\biggl ( \frac{r}{R_{*}} \biggr )^{-\alpha} \;, \label{eq: initial rho} \\
P&=&P_{0}\biggl ( \frac{r}{R_{*}} \biggr )^{-(\alpha+1)} \;, \label{eq: initial P}
\end{eqnarray}
where
\begin{eqnarray}
\alpha & \equiv & \frac{G\rho_{0}M_{*}}{P_{0}R_{*}}-1 \; . \label{eq: alpha}
\end{eqnarray}
Here $\rho_0$ and $P_0$ are the reference density and pressure of the surrounding gas 
on the central star, and $M_*$ is the central stellar mass. The parameter $\alpha $ 
denotes the power-law index of the initial density profile. 

The convectively stable condition for the surrounding gas is 
\begin{equation}
\frac{ds}{dr} > 0 \; ,
\end{equation}
where $s$ is the entropy (Kippenhahn \& Weigert 1990). When adopting equations 
(12)--(15), this can be reduced to 
\begin{equation}
\alpha > \frac{1}{\gamma -1} =3 \;.
\end{equation}
We should set the parameter $\alpha $ to be larger than $3$ in our calculations
for maintaining the surrounding plasma being convectively stable. On the other hand, 
models with $\alpha > 7.25$ can not be solved using our simulation code because the 
physical variables become too small to be resolved in the calculation domain. The parameter 
$\alpha $ surveyed in the following is thus restricted to the range between $3.0$ and $7.25$.

The ratio between the mass and radius of the central star is a fundamental parameter 
which characterizes the system, and is fixed as $M_* /R_* = 2.2\times 10^{27}\ {\rm g/cm}$. 
This is the typical value for the neutron star. All the numerical models we examined 
here are controlled by two non-dimensional parameters, which are the power-law index 
$\alpha $ of the hydrostatic surrounding medium and the plasma beta on the equatorial 
surface of the central star $\beta_0 \equiv P_0/(B_0^2/8\pi )$.

As we show in \S~3.2, the temporal evolution of the magnetically driven outflow is 
controlled by the initial Alf\'ven velocity ($v_{\rm A}$). The larger value of initial plasma beta 
($\propto v_{\rm A}^{-2}$) provides the slower evolution of the outflow, which is not 
appropriate for extensive parameter survey. In addition, the conservative MHD scheme used 
in our code is not good at solving the problem with small values of the initial plasma beta. 
Hence we survey the $\beta_0$-dependence of the outflow properties within the range 
$3.2 < \beta_0 < 3.2\times 10^3$.

The relativistic representations of the sound speed and Alfv\'en velocity $v_{\rm s,0}^{\; R} $ 
and $v_{\rm A,0}^{\; R}$ are, at the equatorial surface of the central star, 
\begin{eqnarray}
v_{\rm s,0}^{\; R}   & = & 
\frac{v_{\rm s,0}}{\sqrt{ 1+ \xi/(\alpha +1)}} \nonumber \\
& \simeq & v_{\rm s,0} \; \biggl ( \because \ \  \xi \equiv \frac{\gamma}{\gamma - 1}\frac{GM_{\rm *}}{R_{\rm *}c^2} \simeq 0.65 \biggr ) \;,  
\\
v_{\rm A,0}^{\; R} & = & \frac{v_{\rm A,0}}{\sqrt{1+[1+ 2(\gamma - 1)/(\gamma \beta_{0})] [\xi/(\alpha + 1)]}} \nonumber \\
& \simeq & v_{\rm A,0} \;, 
\end{eqnarray}
where
\begin{equation}
v_{\rm s,0}   =  \sqrt{\frac{\gamma P_0}{\rho_0}} \;, \ \ \ \ \ v_{\rm A,0}   =  \frac{B_0}{\sqrt{4\pi \rho_0 }}  \;.
\end{equation}
Here $v_{s,0}$ and $v_{{\rm A},0}$ are the sound speed and Alfv\'en velocity at the 
equatorial surface of the central star in the non-relativistic limit. Since the parameter $\alpha$ 
varies from $3.0$ to $7.25$ and $\beta_0$ varies from $3.2$ to $3.2 \times 10^3$ in our 
models, the relativistic representations of the sound and Alfv\'en velocities can be well 
approximated by their non-relativistic forms as stated above. Using description (\ref{eq: alpha}), 
the sound and Alfv\'en velocities are 
\begin{eqnarray}
v_{\rm s,0}/c  & = & 0.48(\alpha + 1)^{-1/2} \;, \label{eq: initial v_s} \label{eq: initial v_s} \\ 
v_{\rm A,0}/c  & = & 0.32(\alpha + 1)^{-1/2} \biggl ( \frac{\beta_0}{3.2} \biggr )^{-1/2} \;. \label{eq: initial v_A}
\end{eqnarray}

In the following, we mainly use the non-relativistic form of the sound and Alfv\'en velocity for 
convenience. The normalization units in length, velocity and time are chosen for all the models 
as radius of the central star $R_*$, light speed $c$, and the light crossing time $\tau = R_* /c$. 
\subsection{Boundary Condition and Numerical Grid}
The stress-free condition is applied to the outer radial boundary located at $r = 200R_{*}$. 
At the inner radial boundary ($r = R_{*}$), we impose the condition that the physical 
variables except $v_{\phi}$ and $B_{\phi}$ are fixed to be time-independent, that is, 
$\partial / \partial t |_{r=R_*} = 0$. The radial and latitudinal velocities of the gas
$v_r$ and $v_\theta $ vanish at the stellar surface. The gas density, the gas pressure and 
the poloidal field strength at the stellar surface are consistently determined from equations 
(9)-(14). The stress-free condition is adopted at the inner boundary only for the toroidal 
magnetic field $B_{\phi}$.

The shearing motion at the inner boundary regulates the energy injection into the calculation 
domain. On the equatorial surface, we impose a shear flow with the longitudinal velocity 
\begin{eqnarray}
v_{\phi} (R_{*}, \theta, t) = \Delta v_{\phi}(t) \Theta \exp [(1-{\Theta}^4)/4] \;,
\end{eqnarray}
where $\Theta = (\theta - \pi/2)/\Delta \theta_{\rm peak}$ (see Mikic \& Linker 1994). 
$\Delta \theta_{\rm peak}$ controls the latitudinal width within which the shearing motion 
is operated, and is fixed as $\pi/ 9$ in the following. 

We investigate two different energy injection patterns. In both cases, the shearing 
velocity $\Delta v_{\phi}(t)$ increases linearly from zero to $0.1c$ between $t = 0$ 
and $10\tau$ and is held at constant velocity $0.1c$ until $t=20\tau$. After $t = 20\tau $, 
the velocity of the shearing motion is kept constant at $0.1c$ in case A, and is 
terminated suddenly in case B.

The amplitude of the shearing velocity is an important parameter which controls the outflow 
dynamics, and should be widely surveyed. However, we focus only on the typical 
case with the fixed shearing velocity of $0.1c$ in this paper. In this case, the shearing 
motion supplies the magnetic energy which is roughly comparable to the thermal energy 
of the system. The dependence of the outflow dynamics on the initial field strength and 
the density profile of the surrounding gas is our interest in this paper. We will study the 
impact of the shearing velocity on the outflow dynamics in the separate paper. 

The polar angle of the calculation domain is uniformly divided into $\Delta \theta = \pi/90$. 
A nonuniform grid is adopted in the radial direction. The radial grid size $\Delta r$ gradually 
increases from $0.036R_{*}$ (at $r=R_*$) to $0.2 R_{*}$ (at $r = 5 R_{*}$), and is 
fixed as $0.2 R_{*}$ outside $r = 5R_{*}$. The largest grid size $\Delta r = 0.2 R_*$ is 
empirically determined to correctly capture the the sharp relativistic shock which propagates 
through the outer calculation domain. The number of the grid points is set as $1024$ in the 
radial and $90$ in the $\theta$ direction for all the models.

\begin{figure*}
\begin{center}
\scalebox{0.93}{{\includegraphics{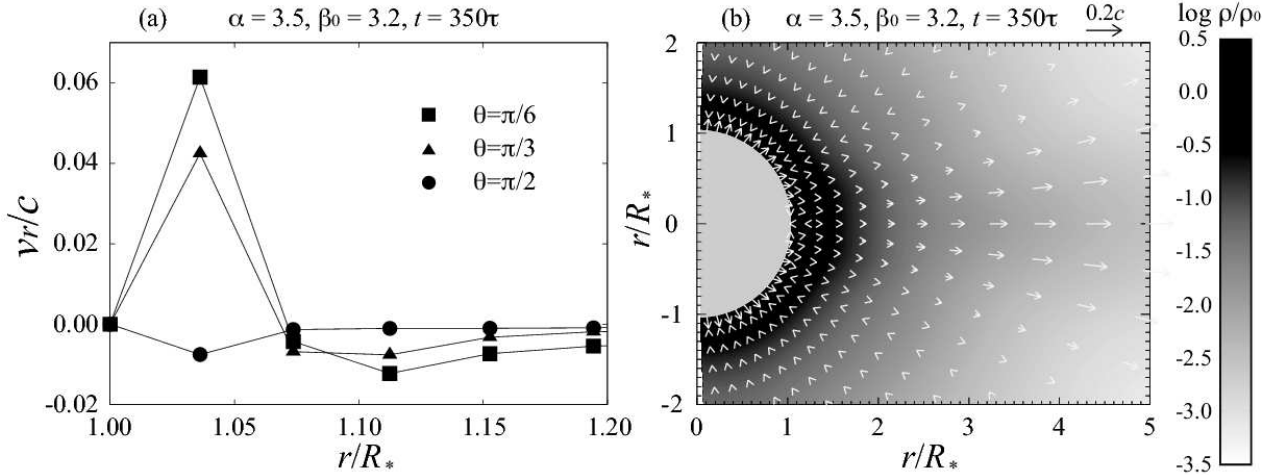}}}
\caption{
Panel (a): the radial profile of the radial velocity for the model where $\alpha=3.5, \beta_0=3.2$ 
at the steady state on around the inner boundary region. The filled square, triangle and circle 
represent the profiles at the different polar angles $\theta=\pi/6$, $\pi/3$, and $\pi/2$ 
(along the equatorial plane), respectively. Panel (b): the density distribution (log scale) and 
poloidal velocity vector near the inner boundary on the meridional plane at the steady state. 
The density is normalized by its inner boundary value. The orientation and length of each 
arrow denote the direction and magnitude of the poloidal velocity vector at each meridian point. 
}
\label{fig1}
\end{center}
\end{figure*}
\begin{figure}
\begin{center}
\scalebox{1}{{\includegraphics{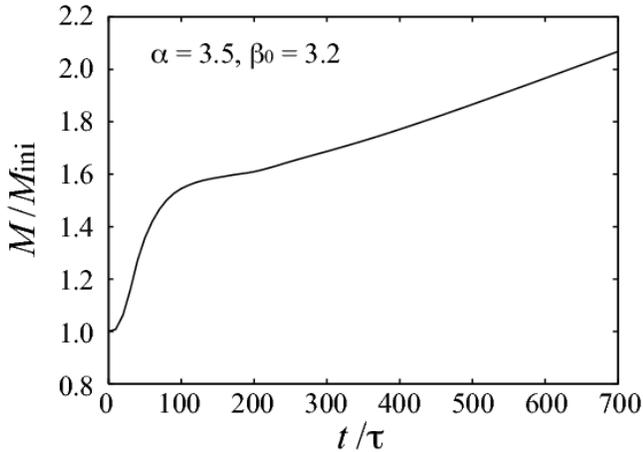}}}
\caption{
The temporal evolution of the total mass contained in the calculation domain for the 
model where $\alpha=3.5, \beta_0=3.2$. The vertical axis is normalized by the initial 
total mass of the plasma distributed in the entire domain.
}
\label{fig2}
\end{center}
\end{figure}

\subsection{Dynamical Effects of the Boundary Condition}
The shearing motion imposed on the equatorial surface of the compact star generates 
the toroidal magnetic field by stretching the initial dipole field. The kinetic energy of 
the shearing motion is then immediately converted to magnetic energy. The magnetic 
energy amplified above the stellar surface initiates the dynamical evolution of the system. 
While the shearing motion is added artificially to break the initial equilibrium of the 
system, it is physically modeling the supply of the magnetic helicity and energy from 
the stellar interior into the magnetosphere, like as the flux emergence in the sun. 

In the solar flare/CME process, the emergence of the tightly twisted magnetic flux plays a 
triggering role in breaking the equilibrium of the corona (Ishii et al. 1998; Kurokawa et al. 2002;  Magara 2003). 
We use the shearing motion instead of the flux emergence for the sun as the process 
modeling energy and helicity injection in our calculations, because actual energy and helicity 
injection processes are still veiled in mystery in the context of compact stars.

Since the inner boundary condition in our models is mathematical over-determined 
(see Bogovalov 1997), there exists a discontinuity in the radial velocity at around 
the stellar surface in our calculations. Figure~\ref{fig1}a depicts the radial profile of the radial velocity 
of the model where $\alpha=3.5$ and $\beta_0=3.2$ as a reference. Here we focus on the 
inner boundary region ($1 \le r/R_{*} \le $1.2) and the time-steady state of $t=350\tau$. The 
filled square, triangle and circle represent the radial velocity profiles at different polar angles 
$\theta=\pi/6$, $\pi/3$, and $\pi/2$ respectively. This addresses that there exists a 
velocity discontinuity near the inner boundary. In addition, the discontinuity 
at around the high latitude stellar surface has a larger amplitude than that around the equatorial 
region where the outflow is mainly accelerated in our models as shown in \S~3. 

Figure~\ref{fig1}b shows the meridional distribution of the density and poloidal velocity near 
the inner boundary at the steady state of the model same as Figure~\ref{fig1}a. 
The orientation and length of each arrow denote the direction and magnitude of the poloidal velocity vector 
at each meridian point. The grey contour is a logarithmic representation of the density
normalized by its inner boundary value. The radial flow across the inner boundary is 
induced especially at around the high latitude stellar surface although the radial velocity is 
set to be zero at the inner boundary. This induced outgoing radial flow supplies the plasma 
into the calculation domain.

The temporal evolution of the total mass contained in the calculation domain is presented in 
Figure~\ref{fig2} in the same model. The vertical axis is normalized by the initial total 
mass of the plasma distributed in the entire domain. The total mass in the system increases with 
time due to the mass inflow from the inner boundary region. The mass injection rate is roughly 
$10^{-2} \; M_{\rm ini}/\tau$ at the initial evolutionary stage until $100\tau$, but 
$10^{-3} \; M_{\rm ini}/\tau$ at the steady state after $200\tau$. 

Since the density of the plasma near the stellar surface is maintained at close to the initial 
value by the mass inflow across the inner boundary as shown in Figure~\ref{fig1}b, the Alf\'ven 
and sound velocity ($\propto \rho^{-1/2}$) 
are not relativistic there and an order of $0.1c$ 
at the steady state. The radial velocity is less than $0.01c$ when focusing on the equatorial 
region where the magnetic energy is strongly amplified and the outflow is energized, 
while its latitudinally-averaged value is roughly $0.05c$.

\begin{figure*}
\begin{center}
\scalebox{1}{\rotatebox{0}{\includegraphics{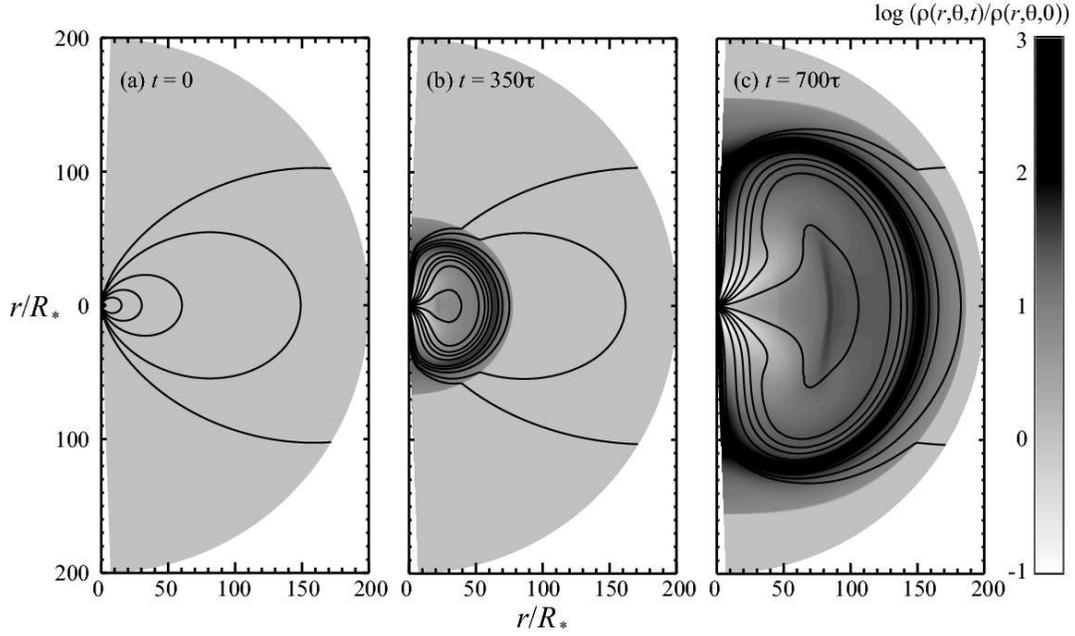}}}
\caption{The time evolution of the density distribution (log scale) and 
magnetic field lines in the meridional plane for the model where 
$\alpha = 3.5$ and $\beta_{0}=3.2$. The density is normalized by its initial 
value: $\rho(r,\theta,t)/\rho(r,\theta,0)$. The left, middle and right panels 
are corresponding to those in $t/\tau = 0$, $350$, and $700$ respectively.}
\label{fig3}
\end{center}
\end{figure*}
\begin{figure*}
\begin{center}
\scalebox{1}{{\includegraphics{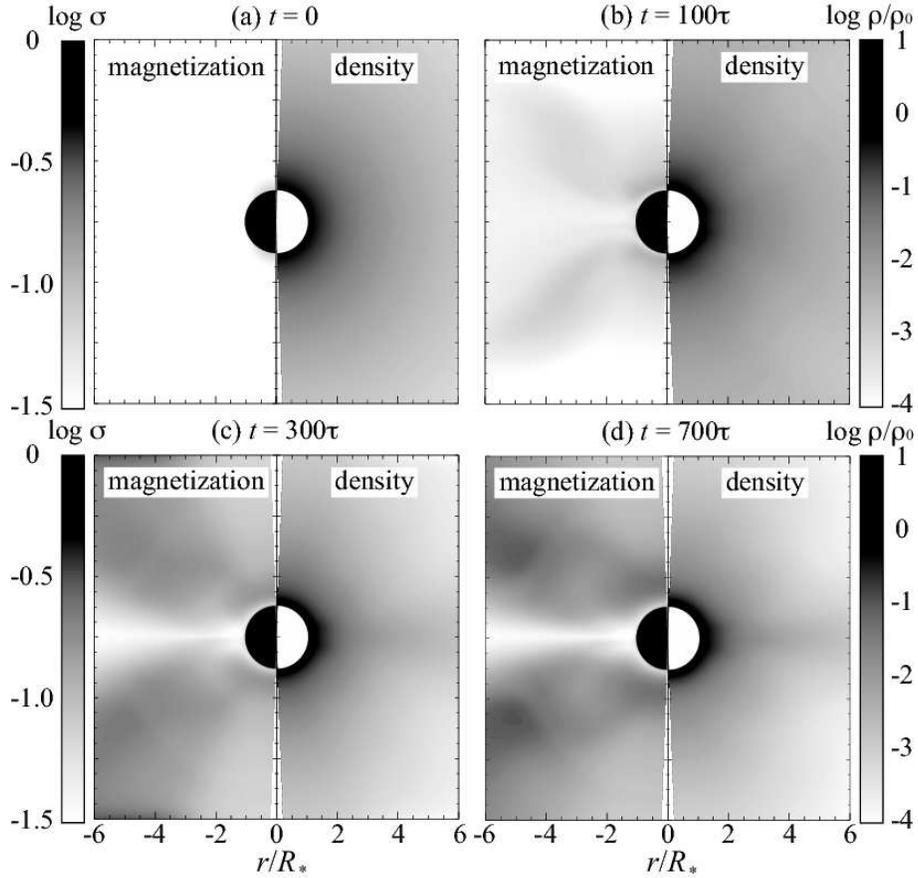}}}
\caption{
The temporal evolution of the magnetization ($\sigma \equiv B^2/4\pi \Gamma^2 \rho c^2$) 
and density on the meridian plane where $\alpha=3.5$ and $\beta_{0}=3.2$ at the times 
(a) $t = 0$, (b) $t = 100\tau$, (c) $t=300\tau$ and (d) $t = 700\tau$ respectively. The density 
is normalized by the reference density of the surrounding gas on the central star. The contours 
are filled by a logarithmic gray scale. The region occupied by the central compact object is filled 
by black and white in the left and right parts of each panel.
}
\label{fig4}
\end{center}
\end{figure*}
\begin{figure}
\begin{center}
\scalebox{1}{{\includegraphics{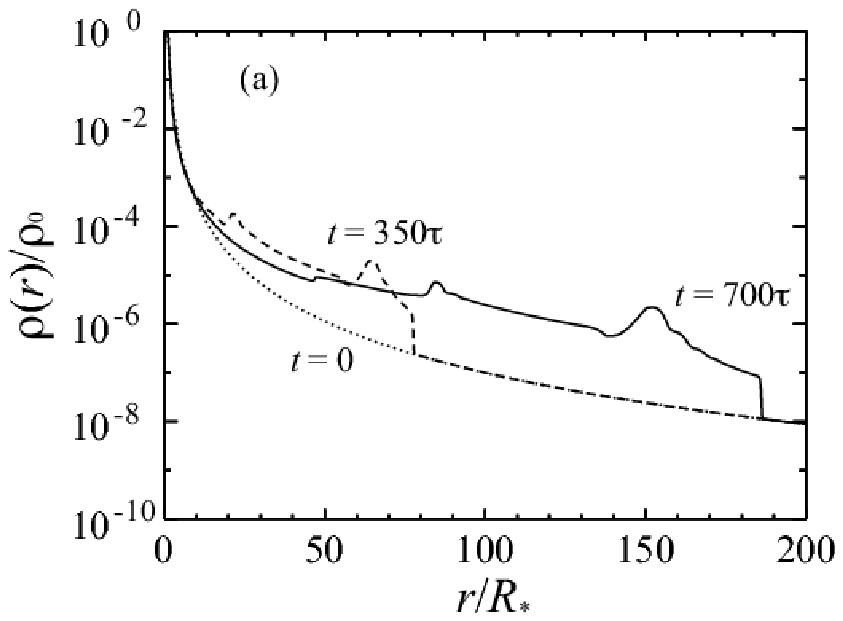}}} \\
\scalebox{1}{{\includegraphics{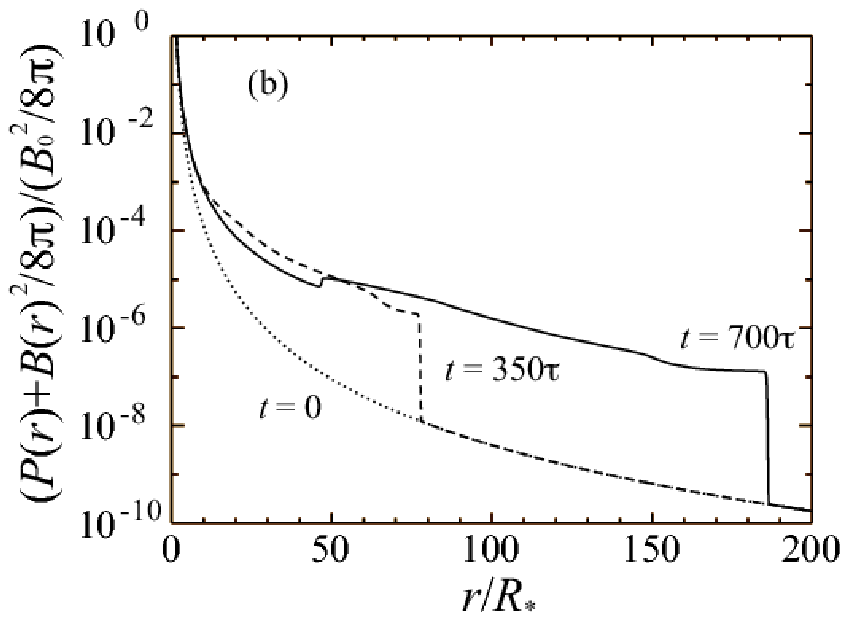}}} \\
\scalebox{1}{{\includegraphics{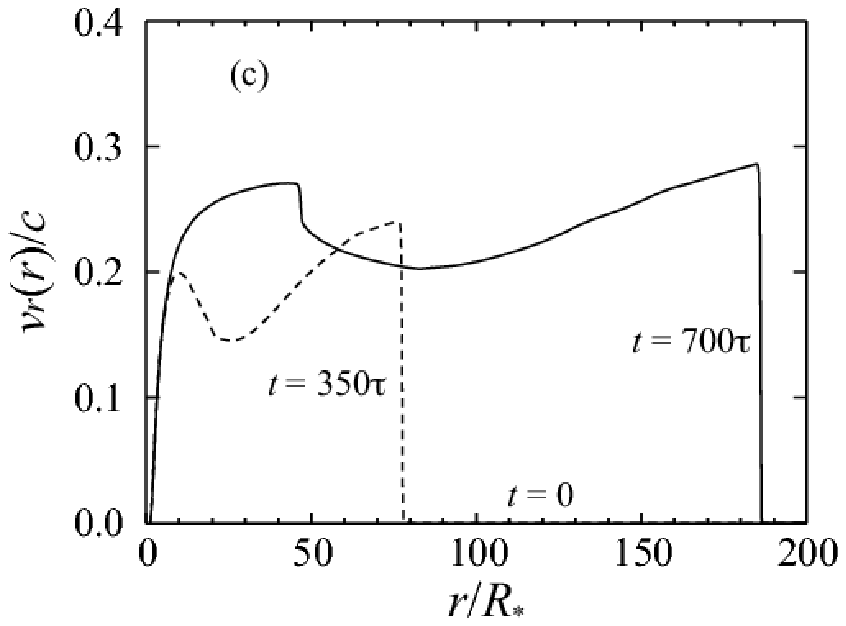}}} 
\caption{Panels (a) the density, (b) the total pressure (thermal + magnetic pressures), 
and (c) the radial velocity on equatorial plane as a function of the normalized radius 
$r/R_{*}$ where $\alpha =3.5$ and $\beta_0=3.2$. Dotted, dashed, and solid lines are 
corresponding to those in t/$\tau$ =0, 350 and 700, respectively.}
\label{fig5}
\end{center}
\end{figure}
\begin{figure*}
\begin{center}
\scalebox{1}{\includegraphics{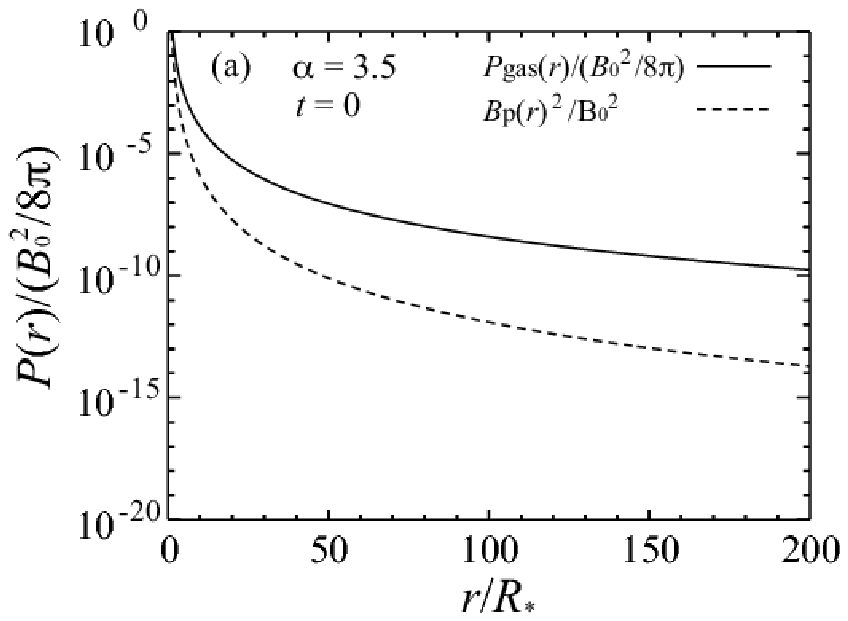}}
\scalebox{1}{\includegraphics{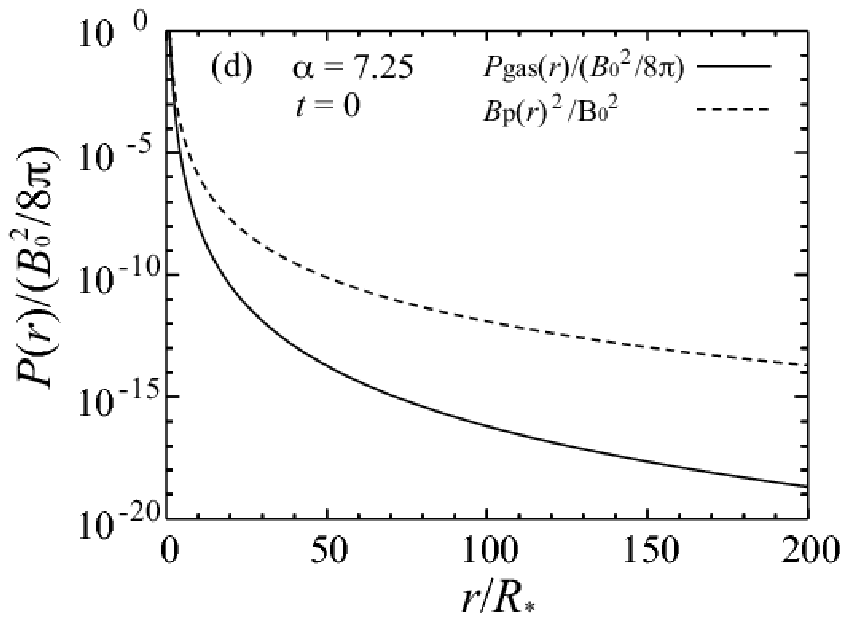}} \\
\scalebox{1}{\includegraphics{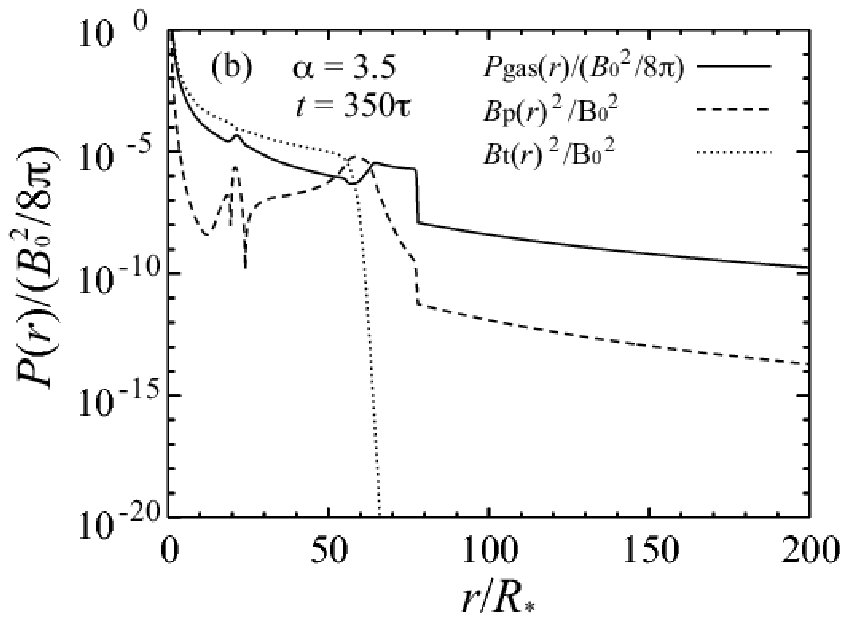}}
\scalebox{1}{\includegraphics{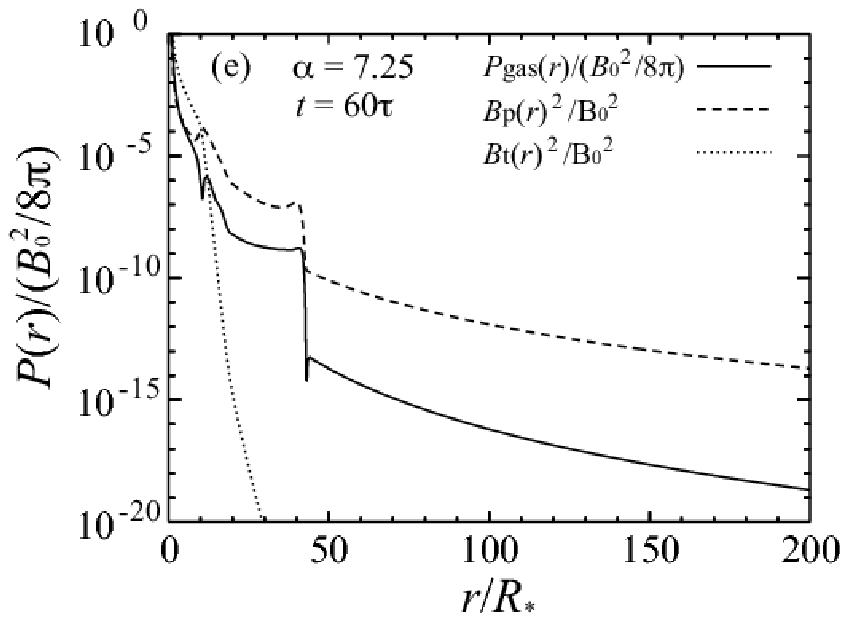}} \\
\scalebox{1}{\includegraphics{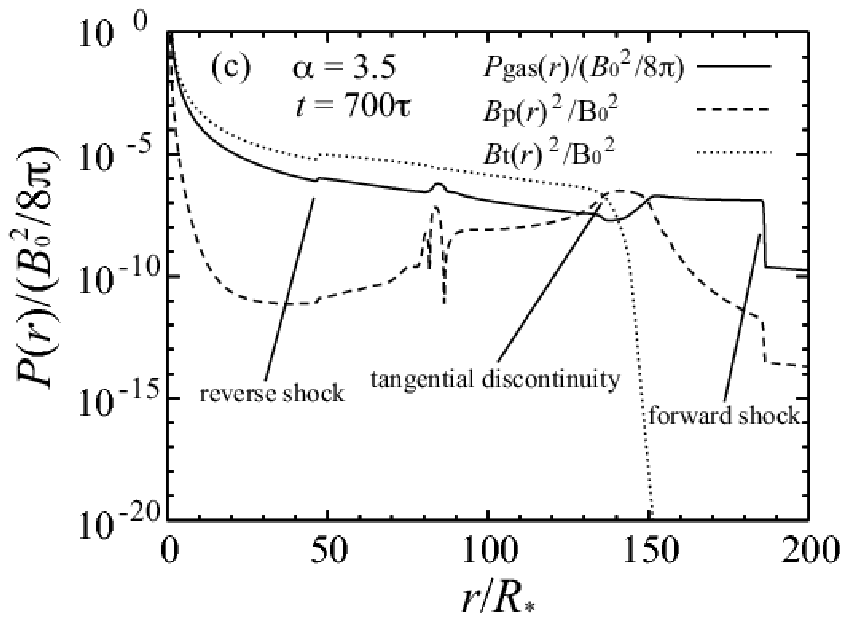}}
\scalebox{1}{\includegraphics{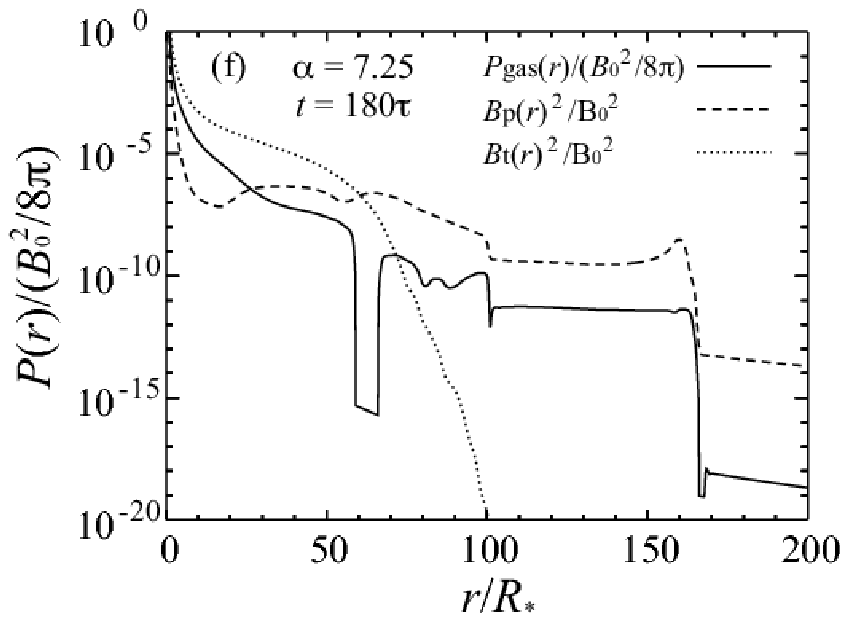}} \\
\caption{Left column: temporal evolution of the radial distribution of the gas 
and magnetic pressures where $\alpha =3.5$ and $\beta_0=3.2$ at the times 
(a) $t=0$, (b) $t = 350\tau$ and (c) $t = 700\tau $ respectively. Right column: 
the model $\alpha =7.25$ and $\beta_0=3.2$. Note that all pressure components 
are measured on the equatorial plane. The magnetic pressure is divided into two parts 
in these panels, one is the contribution from toroidal magnetic fields $B_{t}$, and 
the another is that from the poloidal fields $B_p$.}
\label{fig6}
\end{center}
\end{figure*}
\section{Results}
\subsection{Temporal evolution of a fiducial model}
We show the dynamical evolution of the outflow in a fiducial model ($\alpha = 3.5 $ and 
$ \beta_0 = 3.2 $). The case with continuous shearing motion (case~A) is investigated here.

Figure~\ref{fig3} demonstrates the temporal evolution of the expanding outflow. 
The density and the magnetic field lines projected on the meridional plane are shown 
at the time (a) $t = 0$, (b) $350\tau $ and (c) $700\tau$ respectively. Note that the grey 
contour is a logarithmic representation of the density normalized by its initial value.

The magnetic pressure resulted from the toroidal magnetic field, which is generated by 
the shearing motion around the stellar equatorial surface, initiates and accelerates the 
quasi-spherical outflow. The driven outflow is strongly magnetized, in which the magnetic 
pressure composed mainly of the contribution for the generated toroidal field dominates 
the gas pressure, and carries dense plasma away from the central star.

In the equatorial region where the magnetized outflow is energized, it is natural from 
the physical perspective to consider that the magnetization parameter 
($\sigma \equiv B^2/4\pi \Gamma^2 \rho c^2$) grows due to both the mass ejection 
and the continuous injection of the magnetic energy. However, the magnetization 
parameter near the stellar surface is maintained at a lower value than unity during the computing time
as shown in Figure~\ref{fig4}.

In this figure, the spatial distribution of the magnetization parameter (left) and density 
(right) are demonstrated by the logarithmic grey scale on the meridian plane. Different 
panels are corresponding to the different evolutionary phases (a) $t = 0$, (b) $100\tau$, 
(c) $300\tau$ and (d) $700\tau$ respectively. The region occupied by the central star is 
filled by black and white in the left and right parts of each panel. The magnetization 
$\sigma$ near the stellar surface is less than unity. The mass inflow from the inner 
boundary region plays a role in reducing the magnetization parameter to less than unity. 

Although the magnetic energy is strongly amplified and the outflow carries a large portion 
of the gas around the equatorial region, the rapid amplification of the magnetization parameter 
is prevented because the dense plasma maintained above the high-latitude stellar surface is 
continuously supplied into the equatorial region. This is the reason why the magnetization 
parameter $\sigma $ in the near surface region is less than unity at all the latitude during the 
computing time.

Figure~\ref{fig5} denotes the radial distributions of the (a) density, (b) total pressure 
(thermal + magnetic pressures) and (c) radial velocity along the equatorial plane 
($\theta = \pi/2$). The dotted, dashed and solid lines represent the cases $t = 0$, 
$350\tau$ and $700\tau $ of Figure~\ref{fig3} respectively. Note that the vertical 
axis is normalized by its reference value in each plot.

It is found that a strong forward shock is formed and expands supersonically into the 
surrounding medium. The shock powered by the magnetized outflow is accelerated to 
sub-relativistic velocity $\simeq 0.3c$. Behind the forward shock surface (around 
$50R_{\rm *}$ when $t=700\tau$), a reverse shock is excited and propagates through 
the shocked medium.

The time evolution of the gas and magnetic pressures are shown in Figure~\ref{fig6}. 
The left column is of the case being discussed in this section. The top, middle and bottom 
panels correspond to phases (a) $t = 0$, (b) $350\tau $, and (c) $700\tau $ in the fiducial 
model. In each panel, the solid line indicates the gas pressure, dotted and dashed lines 
are the magnetic pressure due to the poloidal and toroidal magnetic field respectively. 
The normalization of each curve is the initial magnetic pressure at the equatorial surface 
of the central star, that is ${B_0}^2/8\pi$.

In Figure~\ref{fig6}c, we find a tangential discontinuity created behind the propagating 
forward shock. The magnetic pressure provided by the toroidal magnetic field becomes 
predominant behind the discontinuity. In contrast, the gas pressure dominates the magnetic 
pressure between the discontinuity and the shock surface throughout the evolution of 
the system, except for a transition region near the discontinuity.
\subsection{Magnetically Driven Outflow}
The physical properties of the magnetically driven outflow are examined through
changing the $\alpha$-parameter that controls the density profile. We anticipate that 
the magnetically driven outflow velocity $v_{\rm mag}$ is characterized by the 
strength of the initial dipole field which is the seed of the predominantly toroidal field 
behind the tangential discontinuity. The relation between the velocity of the magnetically 
driven outflow and the initial field strength is elucidated here. 

\begin{figure}
\begin{center}
\scalebox{1}{\rotatebox{0}{\includegraphics{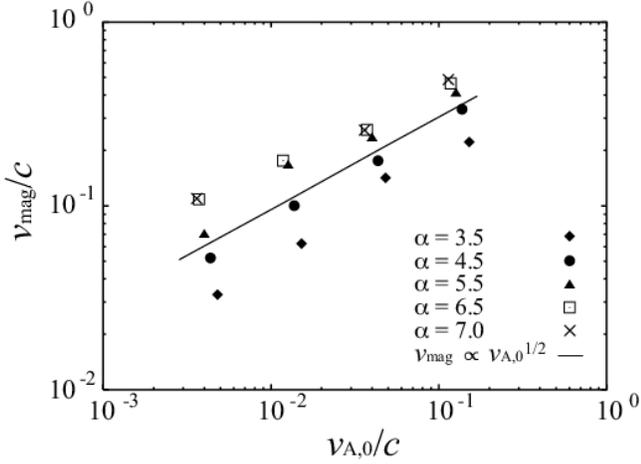}}} \\
\caption{The relation between the Alfv\'en velocity at the equatorial 
surface and the velocity of the tangential discontinuity measured 
at $r=70R_{*}$ when $\alpha = 3.5$, $4.5$, $5.5$, $6.5$, and $7.0$.
There is a simple scaling relation between these two values,
$v_{\rm mag} \propto {v_{\rm A,0}}^{1/2}$.}
\label{fig7}
\end{center}
\end{figure}

Figure~\ref{fig7} shows the outflow velocity $v_{\rm mag}$ in 
relation to the initial Alfv\'en velocity at the equatorial surface of the 
central star ($v_{\rm A,0}$) for the models $\alpha = 3.5$, $4.5$, $5.5$, $6.5$, and $7.0$ respectively. 
The initial Alfv\'en velocity is the representative of the seed field strength and 
varies from $1.1 \times 10^{-3}c$ to $0.15c$ when changing the size of $\alpha$ and $\beta$. 
The velocity is measured when the head of the magnetically driven outflow reaches 
$r = 70R_{*}$. 

The velocity of the magnetically driven outflow increases with the initial Alfv\'en velocity. 
The logarithmic fitting of the data provides the power law 
dependence between them as summarized in Table~1, where $\sigma \equiv {\rm d}\ln v_{\rm mag}/{\rm d}\ln v_{\rm A,0}$
denotes the power law index. The index $\sigma $ is nearly constant for all the models, 
that is $\sigma \simeq 0.5$. The dynamical evolution of the magnetically driven outflow thus follows 
a simple scaling relation, $v_{\rm mag} \propto {v_{\rm A,0}}^{1/2}$.

\clearpage

\begin{table}[!htbp]
\begin{center}
\caption{Indices of a simple relation between the magnetically driven outflow 
velocity at tangential discontinuity surface and the initial Alfv\'en velocity at the 
equatorial surface of the central star when the power law index $\alpha$ of the 
initial density profile varies. The relation is $v_{\rm mag} \propto {v_{\rm A,0}}^{\sigma}$.}
\scalebox{1}{\rotatebox{0}{
\begin{tabular}{@{}cccccc}\hline\hline
$ \alpha $ & 3.5 & 4.5 & 5.5 & 6.5 & 7.0 \\ 
\hline
$ \sigma $ & 0.51 & 0.54 & 0.45 & 0.44 & 0.48 \\ 
\hline\hline
\end{tabular}}}
\label{table1}
\end{center}
\end{table}

To draw a qualitative physical picture which is accountable for this scaling relation, we 
focus on two balancing equations. For simplicity, we reduce the governing equations 
(1)--(8) to their non-relativistic forms because the outflow velocity of interest  here is 
sufficiently slower than the light speed. The numerical results address that in the steady 
state, the strength of the toroidal field behind the tangential discontinuity does not change 
significantly in time. Hence the radial advective loss of the toroidal field should approximately 
counterbalance the generation of the toroidal field by shearing motion at around the stellar surface. 
The azimuthal component of the induction equation can be simplified, at the steady state, 
\begin{eqnarray}
&&\frac{\partial B_{\phi}}{\partial t} = [\nabla \times (\bi{v} \times \bi{B})]_{\phi} \nonumber \\
             \leftrightarrow    &&            0         \simeq  \frac{B_0 \Delta v_{\phi}}{L} - \frac{B_{\phi \rm ,amp} v_{r \rm ,adv}}{R_{*}} 
                             \;, \label{eq: induction}
\end{eqnarray}
where $B_{\phi, \rm amp}$ is the typical strength of the toroidal field in the region 
where it is strongly amplified and $v_{r, \rm adv}$ is the mean radial advection 
velocity there. The first term on the right hand side of equation~(\ref{eq: induction}) 
represents the generation of the toroidal field by the shearing motion within typical 
latitudinal width $L$, which is replaced using $L \simeq R_{*}\Delta \theta_{\rm peak}$ 
in the following. The second term shows the advective loss of the toroidal field with 
the mean radial velocity $v_{r,\rm adv}$. This yields the mean toroidal field in the steady state, 
\begin{eqnarray}
B_{\phi \rm , amp} \simeq \biggl ( \frac{1}{\Delta \theta_{\rm peak}} \biggr ) \biggl ( \frac{\Delta v_{\phi}}{v_{r, \rm adv}} \biggr ) B_0 \; . \label{eq: B_phi,amp}
\end{eqnarray}

Our numerical results suggest that the magnetically driven outflow is mainly accelerated 
in the near-surface region where the toroidal field is strongly amplified as shown in 
Figure~\ref{fig6}c and ~\ref{fig7}c. The velocity of the magnetized outflow does not 
change drastically in the region beyond the radius $r \simeq 10 R_{*}$. This would 
address that large portion of the kinetic energy of the outflow is gained through the 
conversion of the magnetic energy in the near-surface region of the compact star. 
 
In the steady state, the total energy of the magnetically driven outflow should be conserved 
along the streamline, which is almost radial in our models. The specific kinetic energy of 
the outflow at an arbitrary radius, at which the energy conversion is almost terminated, is 
thus given by integrating the magnetic pressure gradient force along the radial streamline 
from the stellar surface $R_{*}$ to the arbitrary radius $r$, 
\begin{eqnarray}
\frac{1}{2}{v_{r}(r)}^2 = - \int^{r}_{R_{*}} \frac{1}{\rho}\frac{\partial}{\partial r} \biggl ( \frac{{B_{\phi}}^2}{8\pi} \biggr ) \; {\rm d}r \label{eq: v_r} \;,
\end{eqnarray}
(see Appendix for details). Since the magnetized outflow is not accelerated dramatically 
after the termination of the energy conversion, its mean radial velocity $v_{\rm mag}$ 
can be substituted for the local value $v_r(r)$ in equation (\ref{eq: v_r}). This energy 
balance equation yields the typical velocity of the magnetized outflow as a function of 
the mean toroidal field $B_{\phi,{\rm amp}}$ and the mean density $\rho_{\rm amp}$ 
in the region where the toroidal field is strongly amplified and the outflow gains the 
kinetic energy, 
\begin{eqnarray}
 {v_{\rm mag}} \simeq \frac{B_{\phi, \rm amp}}{\sqrt{4\pi\rho_{\rm amp}}}\;. \label{eq: energetics}
\end{eqnarray}
As the outflow velocity does not change drastically in the region beyond the near-surface 
of the central star, the mean advection velocity $v_{r,{\rm adv}}$ appeared in equation 
(\ref{eq: B_phi,amp}) should be comparable to the typical velocity of the magnetically 
driven outflow, that is $v_{r, \rm adv}  \simeq v_{\rm mag}. $ Then, by combining 
equations~($\ref{eq: energetics}$) with ($\ref{eq: B_phi,amp}$), we can provide 
\begin{eqnarray}
v_{\rm mag} \simeq {v_{\rm A,0}}^{1/2} {\Delta v_{\phi}}^{1/2} \Delta \theta_{\rm peak} ^{-1/2} 
\left( \frac{\rho_{\rm amp}}{\rho_0} \right )^{-1/4} \;, \label{eq: v_mag scaling}
\end{eqnarray}
where the parameters $v_{\rm A,0}$, $\Delta v_\phi$, and $\Delta \theta_{\rm peak}$ are 
initially given and remained to be constants. The relation (\ref{eq: v_mag scaling}) obtained  
using an order of magnitude analysis here is shown by the solid line in Figure~\ref{fig7}. 
The numerical results are well reproduced by this analytic scaling. It is noted again that two 
balancing equations, which are reduced from the induction and energy equations, account for 
the scaling relation.

The physical property of the magnetized outflow studied here is different from that 
found in the similar work on solar CME by Mikic \& Linker (1994). This would be because, 
in their work, the shearing motion which models the energy injection in the solar flare/CME 
process is much slower than the Alfv\'en velocity in comparison with our models. 
The magnetic energy is gradually stored above the solar surface and is released abruptly 
in their CME models. In contrast, in our models, the magnetic energy is immediately 
converted to the kinetic energy without being stored. Although there is less information 
about the injection mechanism and supply rate of the magnetic energy in the compact star, 
we expect that the rapid injection of the magnetic energy into the magnetosphere might be 
the origin of giant flares and the associated mass ejection in the magnetar system. 
We apply our numerical model to the magnetar system in \S~5.
\subsection{Relativistic Self-Similar Shock}
The second component of the outflow is the shock driven outflow which proceeds
the magnetically driven component. We examine how the initial density profile 
of the surrounding gas affects the shock driven outflow. The plasma beta at the 
equatorial surface is fixed as $\beta_0 = 3.2$ here for simplicity.

\begin{figure}
\begin{center}
\scalebox{0.99}{\rotatebox{0}{\includegraphics{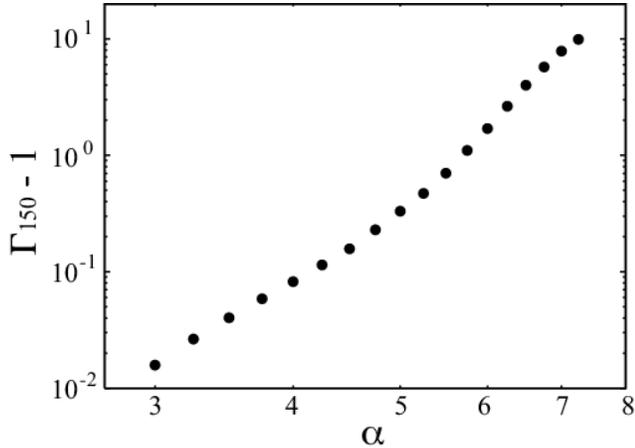}}} \\
\caption{The relation between the initial density profile of the surrounding medium 
($\rho \propto r^{-\alpha}$) and the plasma velocity at the forward shock 
surface on the equatorial plane. Note that the velocity is measured when 
the shock surface reaches the equatorial radius $r = 150 R_{*}$. The vertical 
axis represents $\Gamma_{150} -1 \equiv [1-(v_{150}/c)^2]^{-1/2} -1$ where 
$\Gamma_{150}$ and $v_{150}$ are the Lorentz factor and the plasma velocity 
at the shock surface when it reaches the fixed radius $r = 150 R_{*}$.}
\label{fig8}
\end{center}
\end{figure}

Figure~\ref{fig8} shows the plasma velocity of the shock surface measured at the
reference point $r = 150R_{*}$ and $\theta = \pi/2$ as a function of the index $\alpha$ 
which controls the density profile. The vertical axis denotes $ \Gamma_{150} - 1$, 
where  $\Gamma_{150}$ is the Lorentz factor defined by $[1 - (v_{150} /c)^2]]^{-1/2}$, 
and $v_{150}$ is the plasma velocity of the shock at the reference point. The value 
$\Gamma -1$ measured in this figure is useful for treating both relativistic and 
non-relativistic flows simultaneously. When the flow velocity reaches the ultra-relativistic 
regime, it roughly provides the Lorentz factor itself, that is $\Gamma -1 \simeq \Gamma$. 
In contrast, it is the indicator of the flow velocity also in the non-relativistic case $v \ll c$, 
that is $\Gamma -1 \simeq (v/c)^2/2$. It is found that the shocked plasma is eventually 
accelerated to mildly relativistic velocity in the model $\alpha = 7.25$. There is a correlation 
between the shock velocity and the index $\alpha$. The $\alpha$-dependence of the Lorentz 
factor changes at $\alpha \simeq 5$. These indicate that the density profile of the surrounding 
gas is key to reveal the acceleration mechanism of the forward shock in our model.

The right column in Figure~\ref{fig6} depicts the radial profile of pressure components 
along the equatorial plane at the time $t= 0$, $60\tau$ and $180\tau$ for the model 
with the steep density profile ($\alpha = 7.25$). The curves in panels~(e), (f) and (g) 
have the same meaning as those for the model $\alpha = 3.5$. The toroidal magnetic field 
generated by the shearing motion initially powers the magnetized component of the outflow. 
The forward shock is then excited and propagates through the surrounding medium in 
the same way as the model with $\alpha =3.5$.

There is a clear difference between the models with $\alpha = 3.5$ and $\alpha = 7.25$. 
That is the radial structure of the poloidal magnetic field. In the model with $\alpha = 3.5$, 
the poloidal field is accumulated in front of the tangential discontinuity. The radial 
pressure gradient of the accumulated poloidal field seems to push the shock outward. 
This suggests that the shock wave is mainly accelerated by the magnetic pressure gradient 
force when the density profile moderately declines ($\alpha \lesssim 5$).

In contrast, there is no steep gradient in the poloidal magnetic field behind the shock 
in the model $\alpha =7.25$. Rather the propagation surface of the poloidal field 
coincides with the shock surface. This suggests that the excited shock wave propagates 
with the accumulated poloidal field when the density profile declines steeply ($\alpha \gtrsim 5$).

\begin{figure}
\begin{center}
\scalebox{0.99}{\rotatebox{0}{\includegraphics{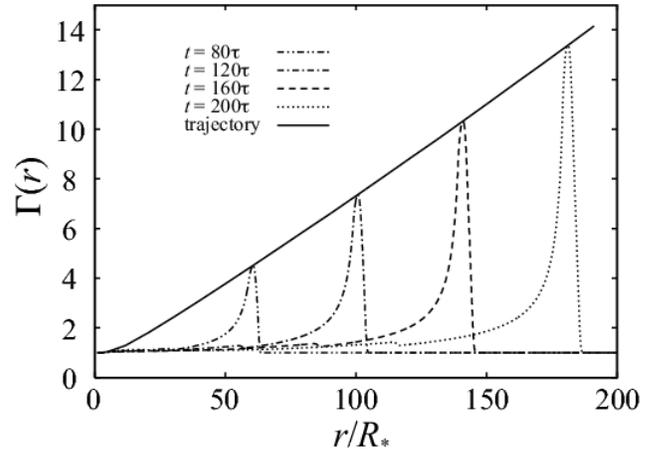}}}
\caption{The time evolution of the Lorentz factor of the expanding gas 
in the model with $\alpha = 7.25$ and $\beta_0=3.2$. The dashed-two dotted,  dot-dashed, 
dashed, and dotted curves indicate the models when $t/\tau=80$, $120$, $160$, and $200$ 
respectively. The solid line represents the time trajectory of the Lorentz factor of the fluid 
velocity at the shock surface measured in the laboratory frame. The scaling relation between 
the Lorentz factor $\Gamma_{\rm sh}$ and spherical radius $r_{\rm sh}$ of the shock surface 
in the laboratory frame is $\Gamma_{\rm sh} \propto r_{\rm sh}$. The relativistic shock 
evolves self-similarly.}
\label{fig9}
\end{center}
\end{figure}

Temporal evolution of the radial velocity profile in the model $\alpha = 7.25$ is presented 
along the equatorial plane in Figure~\ref{fig9}. The vertical axis shows the Lorentz factor 
of the expanding gas. The horizontal axis is the equatorial radius. The dotted, dashed, 
dash-dotted, and dashed-two dotted curves correspond to the cases $t/\tau =80$, $120$, 
$160$ and $200$ respectively. The solid curve denotes the time trajectory of the Lorentz 
factor of the plasma velocity measured at the shock surface in the laboratory frame. 
The evolution of the relativistic shock has self-similar properties. The fitting of numerical 
data provides a scaling relation between the Lorentz factor $\Gamma_{\rm sh}$ and the 
equatorial radius $r_{\rm sh}$ at the shock surface 
\begin{equation}
\Gamma_{\rm sh} \propto r_{\rm sh} \;. \label{eq: self-similar}
\end{equation}
The Lorentz factor of the plasma velocity at the shock surface increases linearly with 
the equatorial radius.

\begin{table}
\begin{center}
\caption{
Indices of a relation between the Lorentz factor $\Gamma_{\rm sh}$ and 
the equatorial radius $r_{\rm sh}$ at the shock surface in the laboratory frame 
when the power law index $\alpha$ of the initial density profile varies.
The relation is $\Gamma_{\rm sh} - 1 \propto {r_{\rm sh}}^{\delta}$.
}
\scalebox{1}{\rotatebox{0}{
\begin{tabular}{@{}cccccccc}\hline\hline
$ \alpha $ & 5.75 & 6.00 & 6.25 & 6.50 & 6.75 & 7.00 & 7.25 \\ 
\hline
$ \delta $ & 0.99 & 1.03 & 1.05 & 1.08 & 1.10 & 1.11 & 1.10 \\ 
\hline\hline
\end{tabular}}}
\label{table2}
\end{center}
\end{table}

The other models with steep density gradient of $\alpha > 5.5$ commonly yield a 
relation close to equation (\ref{eq: self-similar}). Table 2 summarizes the power-law 
index $\delta$ when fitting the numerical models by a power-law relation 
$\Gamma_{\rm sh} - 1 \propto {r_{\rm sh}}^\delta$. It is important to stress that all 
the models which follow the scaling relation (\ref{eq: self-similar}) have relativistic 
velocity, that is, $\Gamma_{150} > 2$, at the reference point. The models with 
$\alpha \lesssim 5.5$ do not reach relativistic velocity at the reference point and not 
follow the relation (\ref{eq: self-similar}). In \S~5, the self-similar solution numerically 
discovered in our work is compared to the other analytic solutions, including the 
Blandford-McKee solution (Blandford \& McKee 1976). 

At the initial stage, the magnetic pressure,  gas pressure, and gas density decrease 
with radius according to $r^{-6}$, $r^{-\alpha-1}$, and $r^{-\alpha}$ respectively 
[see equations (\ref{eq: dipole Br})--(\ref{eq: initial P})]. The plasma beta of the 
surrounding gas finally becomes smaller than unity for the model with $\alpha > 5$ at 
a great distance even if it is larger than unity at the stellar surface. Moreover, the Alf\'ven 
velocity of the initial surrounding gas increases with radius only for the model with 
$\alpha > 6$. These imply that a critical point, at which the MHD effect on the outflow 
dynamics will be altered, might exist at around $\alpha =5$--$6$. Though the MHD 
effect might lead to some change in dynamics of the outflow, we will show that pure 
hydrodynamic mechanism can be responsible for the self-similar property of the shock
by using a simplified one-dimensional hydrodynamic simulationl in \S~4.

Our numerical models indicate that the two-component outflow would be a natural 
outcome of the magnetic explosion on a compact star. In the early stages, the magnetically 
driven outflow forms as a consequence of the rapidly rising magnetic pressure around 
the stellar surface. The magnetized outflow loaded with a dense circumstellar plasma 
expands, and finally excites a relativistic forward shock which has self-similar properties. 
Our two-component outflow model is applied to the magnetar system in \S~5.2. 
\subsection{Impact of Energy Injection Pattern}
\begin{figure}
\begin{center}
\scalebox{0.99}{\includegraphics{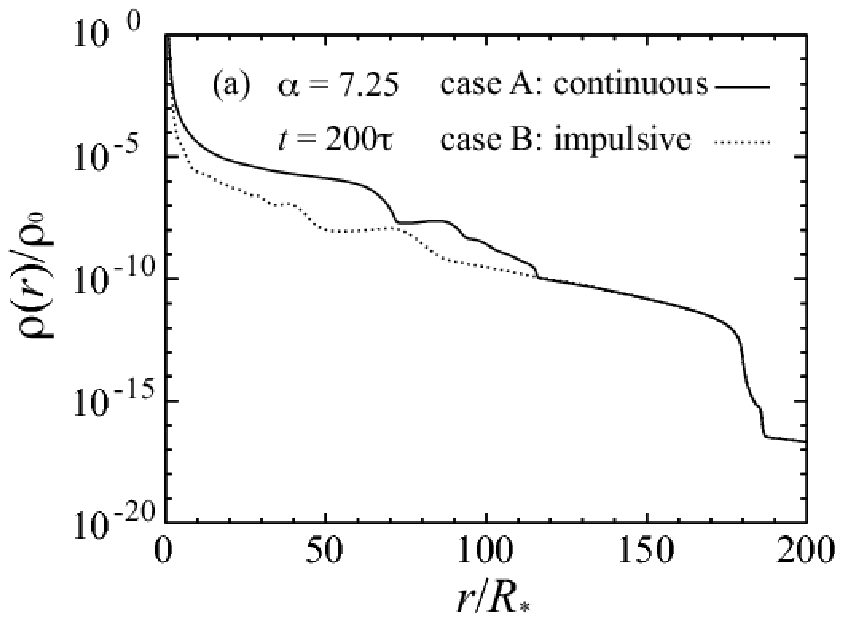}} \\
\scalebox{0.99}{\includegraphics{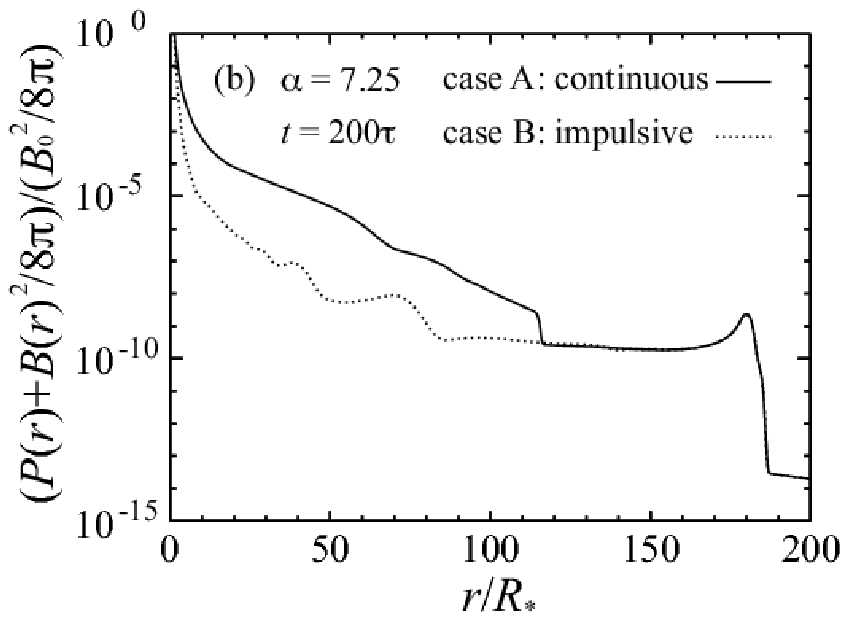}} \\
\scalebox{0.99}{\includegraphics{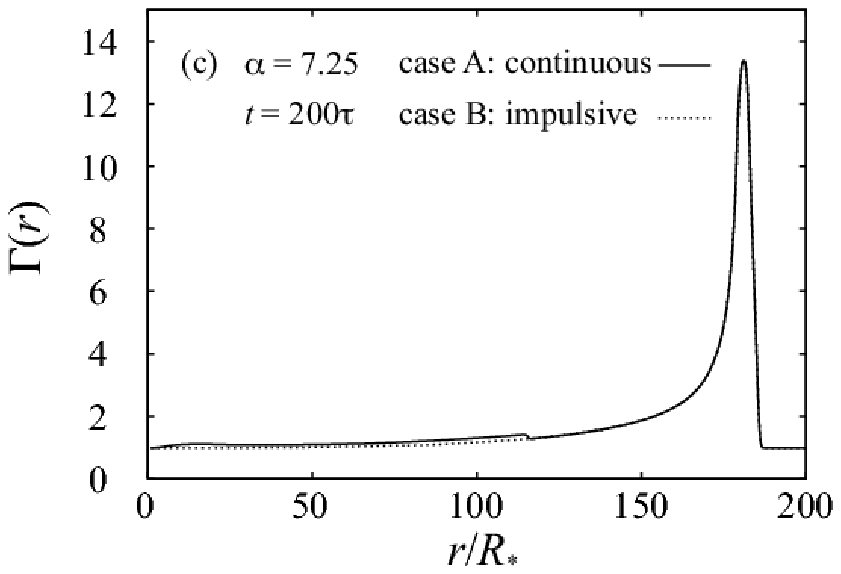}}
\caption{The radial structures of the density, total pressure, and 
Lorentz factor of the plasma as a function of the normalized radius 
$r/R_{*}$ when $\alpha =7.25$, $\beta_0=3.2$, and $t=200\tau$. Case~A, which 
is expressed by the solid line, corresponds to the case with continuous 
energy injection due to the shearing motion. Case~B, which is depicted by 
the dashed line, indicates the case with impulsive energy injection. 
The shearing motion is terminated when $t = 20\tau$ in Case~B.}
\label{fig10}
\end{center}
\end{figure}
\begin{figure}
\begin{center}
\scalebox{1}{\includegraphics{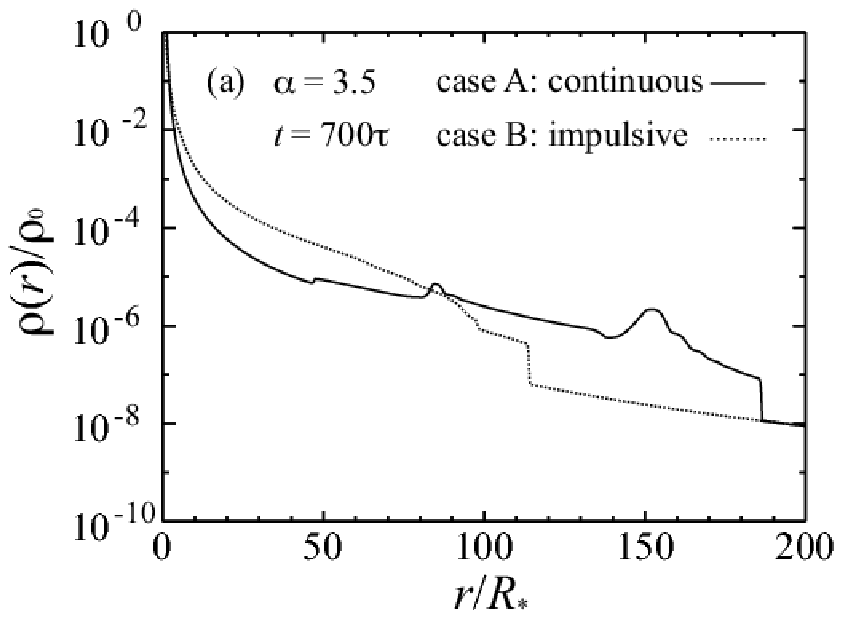}} \\
\scalebox{1}{\includegraphics{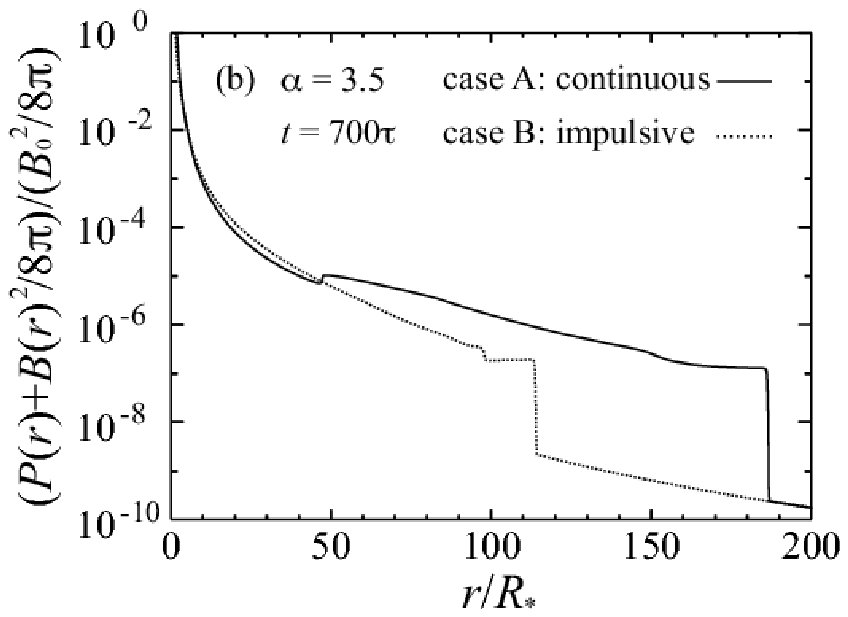}} \\
\scalebox{1}{\includegraphics{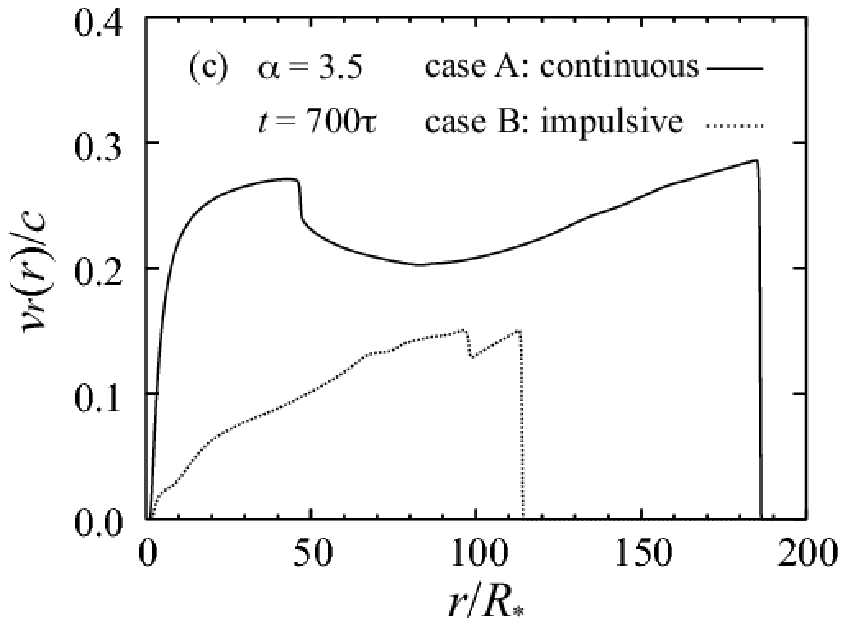}}
\caption{
Almost same as Fig~10, but where $\alpha = 3.5$ and $t=700\tau$.
}
\label{fig11}
\end{center}
\end{figure}
\begin{figure}
\begin{center}
\scalebox{1}{\includegraphics{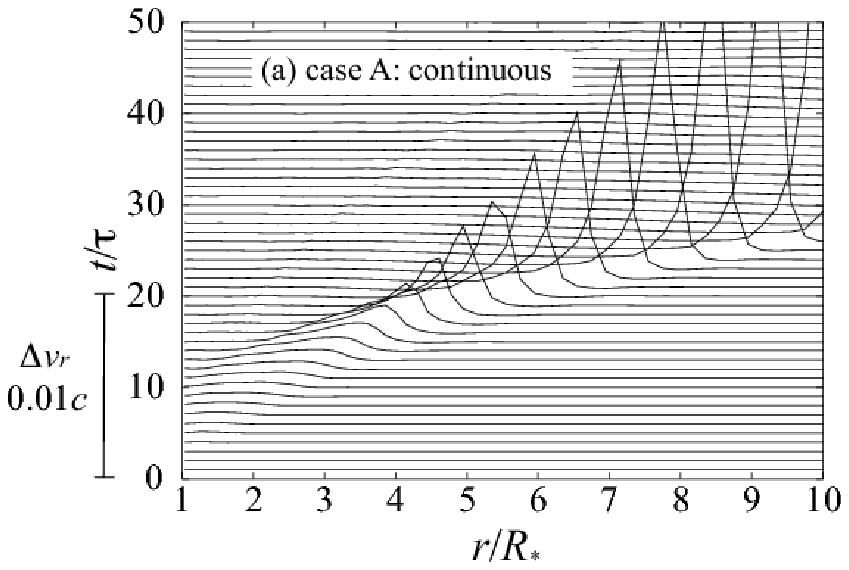}} \\
\scalebox{1}{\includegraphics{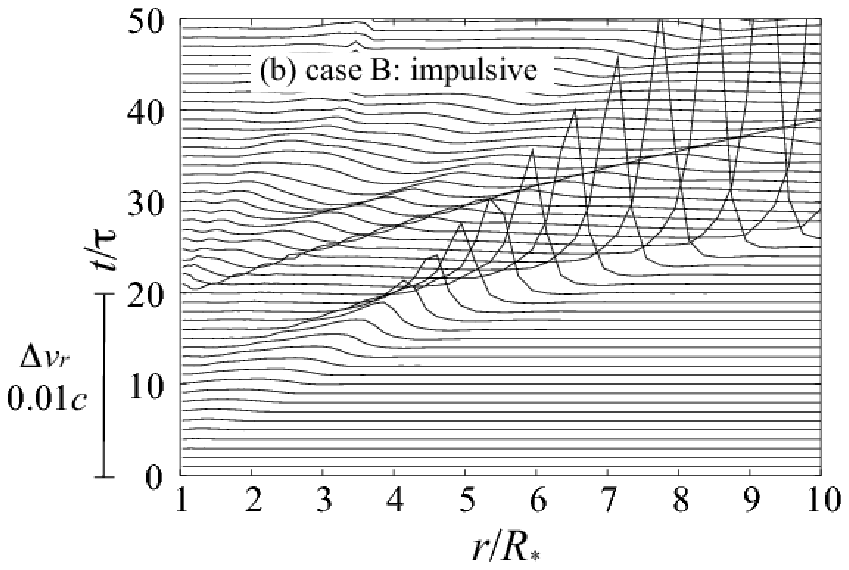}} \\
\scalebox{1}{\includegraphics{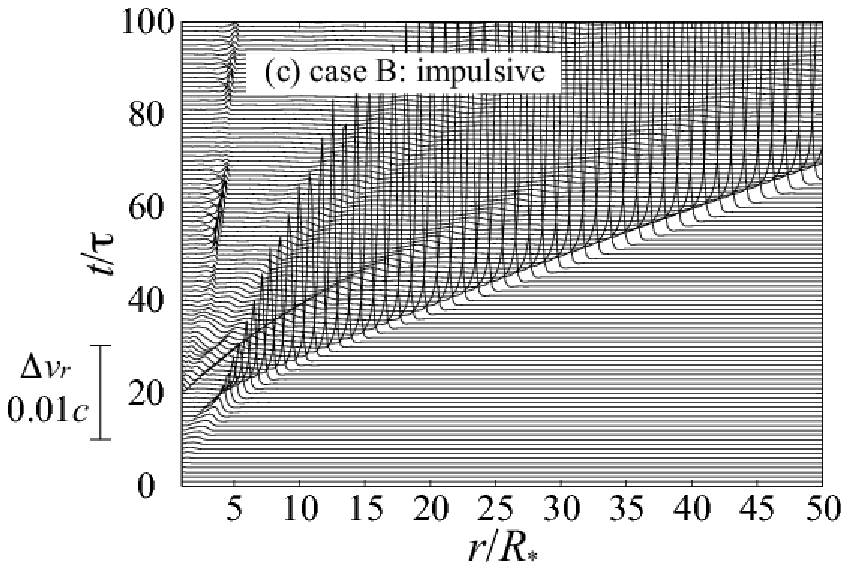}}
\caption{Time-distance diagram for the time difference of the radial velocity 
$\Delta v_r(t,r)$ on the equatorial plane in the model $\alpha =7.25$ where 
$\Delta v_r(t,r) \equiv v_r(t,r) - v_r(t-\Delta t,r)$. Panel (a) corresponds to the 
continuous injection case (Case~A), and (b) is the impulsive injection case (Case~B). 
A wave which transmits the information for the termination of the energy injection is 
excited at $t = 20\tau $ in the case~B. Panel (c) just describes the long-time evolution of 
$\Delta v_r$ in case B. The information wave never reaches the propagating forward shock.}
\label{fig12}
\end{center}
\end{figure}
\begin{figure}
\begin{center}
\scalebox{1}{\includegraphics{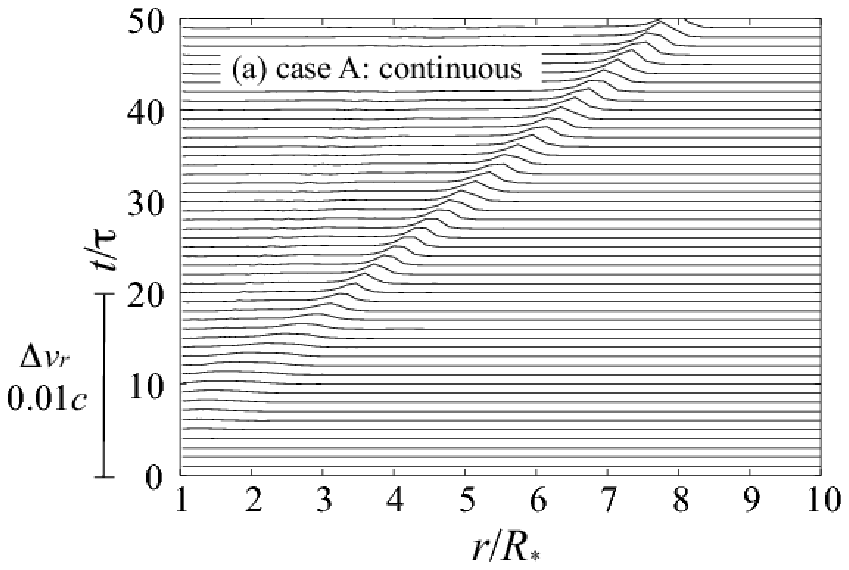}} \\
\scalebox{1}{\includegraphics{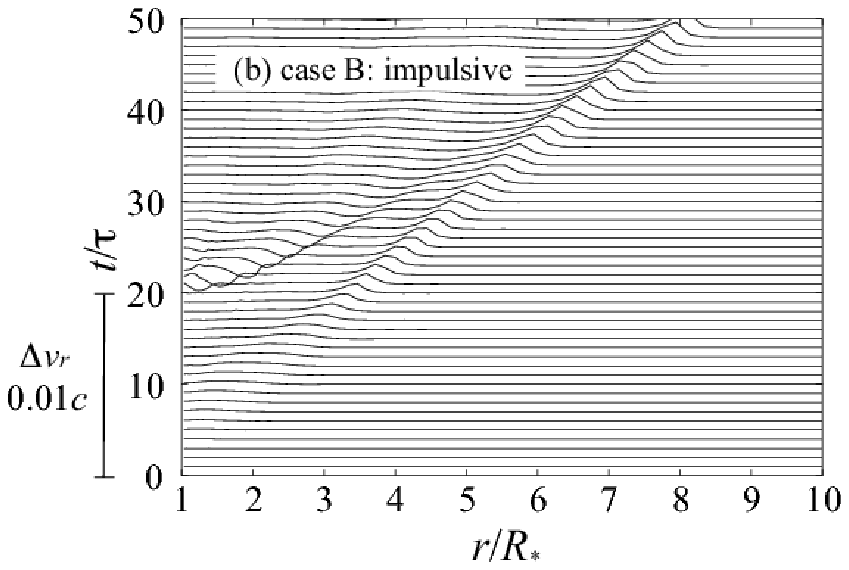}}
\caption{Almost same as Fig~12, but when $\alpha = 3.5$. In this model, 
the information wave can reach the propagating forward shock.}
\label{fig13}
\end{center}
\end{figure}
We clarify the impact of the energy injection pattern at the inner radial boundary on the 
evolution dynamics of the shock. As stated in \S~2, we impose two different kinds of 
the energy injection pattern. These are the continuous injection case with the fixed shearing 
motion (case A) and the impulsive injection case in which we stop the shearing motion 
after $t=20\tau$ (case B). The plasma beta at the equator is fixed again as $\beta_0 = 3.2$.

Figure~\ref{fig10} depicts the radial distribution of the density, total pressure and 
Lorentz factor of the outflow measured at the equatorial plane for both cases when 
$t = 200\tau$. The dashed and solid curves represent the cases~A and B respectively. 
We choose the initial density profile with $\alpha = 7.25$ for this figure. While there 
are differences in the structure of the magnetically driven outflow far behind the shock 
surface, we can not find any difference among two cases around the forward shock.

There is little difference in the structure around the shock between two cases when 
$\alpha > 5.25$. However, for the model with $\alpha < 5.25$, the different injection 
patterns provide different spatial structures of the shock as in the structures of the 
magnetically driven component. This can be verified by Figure~\ref{fig12}, which 
shows the radial distribution of the density, total pressure and velocity in the fiducial 
model ($\alpha =3.5$) when $t=700\tau$ for the case A and case B. The curves have 
the same meanings as those in Figure~\ref{fig11}.

We define a time difference value of the radial velocity on the equatorial plane $\Delta v_r$ as
\begin{equation}
\Delta v_r (t,r) \equiv v_r(t,r) - v_r(t-\Delta t,r) \;,
\end{equation}
where $\Delta t$ denotes the numerical timestep. The time-distance diagram for the time 
difference value $\Delta v_r (t,r)$ is presented for the models $\alpha =7.25$ (Fig.~\ref{fig12}) 
and $\alpha = 3.5$ (Fig.~\ref{fig13}). The vertical and horizontal axes denote the normalized 
time and radius. For specifying the temporal evolution of the outflow, the radial profiles of 
$\Delta v_r$ are plotted at every reference timestep on the time-distance plane. The reference 
timestep is set by the light crossing time $\tau = R_{*}/c$. Panel~(a) corresponds to the case~A 
and panel (b) is the case~B in each figure. 

In the case~B, a wave is excited at $t = 20\tau $ which corresponds to the termination 
time of the energy injection while it is not seen in the case~A. This wave would thus 
transmit the information related to the energy injection pattern to the surrounding gas. 
We call this wave as "information wave" hereafter. The information wave propagates through 
the surrounding medium, but never reaches the forward shock in the model $\alpha = 7.25$ 
(see Figure~\ref{fig12}b). This is because the propagation velocity of the forward shock 
is roughly light speed $c$ while that of the information wave is about $2c/3$.

Figure~\ref{fig12}c is almost same as Figure~\ref{fig12}b, but follows the long-term 
evolution of the time difference value. The excited information wave never interacts with 
the forward shock even when we follow its long-term evolution. This indicates that the inner 
boundary condition does not affect the evolution dynamics of the shock once it has been 
excited when the initial density profile is sufficiently steep ($\alpha > 5.25$).

It would be natural to consider the effect of the termination time of the shearing motion on 
the shock dynamics in the impulsive injection case. There is actually a critical time 
$t_{\rm crit}$ beyond which the termination does not impact on the shock dynamics, that is 
$t_{\rm crit} \simeq 10\tau$. When stopping the shearing motion at the time before $t_{\rm crit} $, 
the information wave can catch up with the shock and change its dynamics even if the initial 
density profile is sufficiently steep. This is because the shearing velocity 
at the inner boundary $\Delta v_\phi (t )$ increases linearly from zero to $0.1c$ during 
$0 \le t \le 10\tau $ (see \S~2.3). The energy injection rate which determines the initial shock 
velocity becomes small if the shearing motion is terminated before $10\tau$. This is the reason 
why there exist the critical time. The shock dynamics does not change even if we stop the shearing 
motion at the time shorter than $20 \tau $ as long as it continues longer than the critical time. 
\section{One-dimensional Hydrodynamic simulation} 
We examine the self-similar properties of the shock driven relativistic outflow in more 
detail using one-dimensional (1D) special relativistic hydrodynamic (HD) simulations, 
which enables us to study the numerical model in a wide parameter range. We specify 
here that a purely hydrodynamic process is responsible for the self-similar properties of 
the relativistic shock discovered in the two-dimensional (2D) MHD models. 

\subsection{Numerical Setting}
We solve the 1D spherically symmetric system with $1024$ grid points radially using 
the same grid spacings as those in the 2D-MHD model. The MHD effects are fully 
ignored by dropping all the related terms in governing equations (1)-(8).

The numerical settings adopted in 1D-HD model are the same as those adopted in the 
2D-MHD model except for the magnetic field. We consider a central compact star 
surround by a hydrostatic gas with the temperature, density, and pressure distributions 
described by equations~(12)--(14). Since the sound velocity follows equation 
(\ref{eq: initial v_s}) with the index $\alpha$ varying from $3.0$ to $7.25$, it becomes 
non-relativistic ($\simeq 0.2c$) at the stellar surface. The symbols used in this section 
have the same meanings as those in the 2D-MHD model (see \S~2.2).

The stress-free condition is applied to the outer boundary located at $r = 200R_{*}$.
At the inner boundary ($r = R_{*}$), we employ the condition that the physical 
variables except velocity are fixed. A plasma inflow is imposed on the inner boundary 
in 1D-HD models as a substitute for the shearing motion which is the energetic origin 
of the outflow in 2D-MHD models. This plasma flow excites the shock-driven outflow 
in the 1D models. The relation between the inflow velocity $v_{\rm in}$ and the injected 
energy flux $F_{\rm in}$ is given by $F_{\rm in} = \rho_0 v_{\rm in}^3/2$. We study 
the properties of the shock driven outflow in the models with the inflow velocities 
$v_{\rm in}/c=1/10$, $1/30$, and $1/100$.  Their dependence on the index $\alpha$ is 
also our interest in this section.
\subsection{Properties of the Shock Driven Outflow}
\begin{figure}
\begin{center}
\scalebox{1}{\rotatebox{0}{\includegraphics{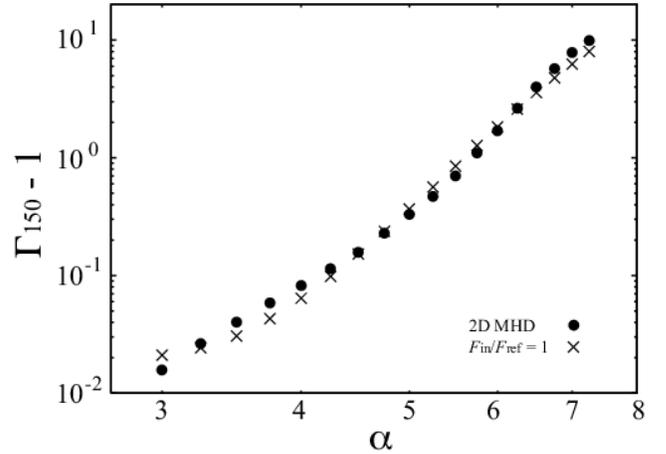}}}
\caption{The Lorentz factor of the shocked plasma velocity at $r = 150R_{*}$
as a function of the density power-law index $\alpha$. The cross denotes 
the model calculated by one dimensional (1D) hydrodynamic simulations. 
The filled circle indicates the 2D special relativistic MHD model shown in Fig.8 again 
for the comparison. The meanings of vertical and horizontal axes are same as Fig.8. 
Note that $F_{\rm ref}$ is a reference value of the inflow energy
flux which corresponds to the case with the inflow velocity $0.1c$.}
\label{fig14}
\end{center}
\end{figure}

The continuous inflow powers a shock driven outflow. The cross symbol in Figure~\ref{fig14} 
shows the plasma velocity of the shock surface measured at $r = 150R_{*}$ as a function 
of the power-law index $\alpha$. Note that the vertical axis is represented by $\Gamma_{150} - 1$ 
like as Figure~\ref{fig8}. As a reference, the 1D-HD model with $F_{\rm in}/F_{\rm ref} = 1$ 
is focused, where $F_{\rm ref}$ is a reference value of the inflow energy flux which is 
equivalent to the model with $v_{\rm in}/c=1/10$. This model provides the almost same physical 
properties of the shock as those observed in the 2D-MHD model with a fixed shearing velocity $0.1c$. 
This is verified by Figure~\ref{fig14}, in which the result of the 2D-MHD model is presented by 
the filled circle.

The $\Gamma_{150}$-$\alpha$ relation in the 1D-HD model can reproduce that in the 
2D-MHD model. The condition for the plasma velocity being relativistic in the 1D-HD 
model is also same as that in the 2D-MHD model, that is $\alpha > 5.5$. These indicate 
that the shock driven outflow is accelerated to relativistic velocity by a purely hydrodynamic 
process even in the 2D-MHD model. 

\begin{figure}
\begin{center}
\scalebox{1}{\rotatebox{0}{\includegraphics{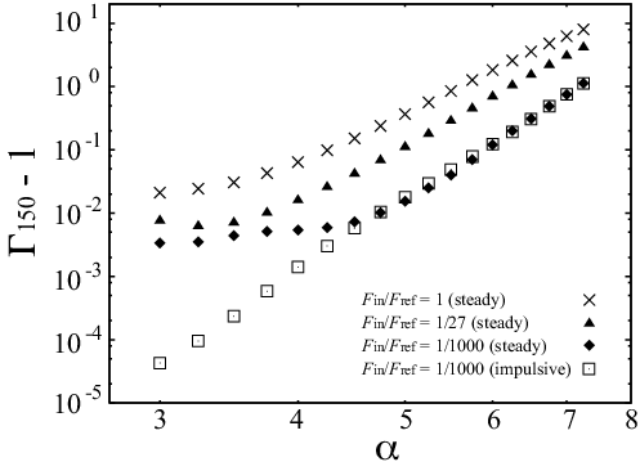}}}
\caption{$\Gamma_{150}$-$\alpha$ relations in the 1D steady 
energy injection model with $F_{\rm in}/F_{\rm ref} = 1$, $1/27$, $1/1000$ 
and impulsive energy injection.}
\label{fig15}
\end{center}
\end{figure}

Figure~\ref{fig15} demonstrates the $\Gamma_{150}$-$\alpha$ relation for the models with 
different inflow energies. The cross, filled triangle, and filled diamond denote the models with 
$F_{\rm in}/F_{\rm ref} = 1$, $1/27$, and $1/1000$ (that is, $v_{\rm in}/c = 1/10$, $1/30$ 
and $1/100$) respectively. The Lorentz factor becomes higher for the model with the larger 
inflow flux. It is important to stress that the condition for the plasma velocity being relativistic 
($\Gamma_{150}>2$) changes depending on the amount of the inflow flux. When setting the 
lower value of the $\alpha $ for providing the relativistic velocity as $\alpha_{\rm trans}$, we can 
find $\alpha_{\rm trans} = 5.5$, $6.0$ and $7.0$ for the model with $F_{\rm in}/F_{\rm ref}=1$, 
$1/27$ and $1/1000$ respectively. The transition value $\alpha_{\rm trans}$, at which the shock 
velocity undergoes transition to the relativistic velocity, depends on the inflow flux. The larger 
inflow flux provides the smaller transition value. 

Note that the transition value $\alpha_{\rm trans}$ depends on the location where we measure 
the plasma velocity. We expect that the shock driven outflow is accelerated and finally reaches 
the relativistic velocity for all the continuous injection models as long as the initial density profile 
has a index $\alpha $ at least larger than $3$ (the smallest value for our parameter survey). With 
fixed inflow flux, the transition value $\alpha_{\rm trans} $ thus is expected to become smaller 
if the plasma velocity of the shock is measured at a more distant location. In other words, transition 
value is a function of the inflow flux and the location where the plasma velocity is measured. 

Separately from the transition value, Figure~\ref{fig15} indicates that there exists a critical 
value of the index $\alpha$, at which the $\alpha$-dependence of the plasma velocity changes. 
The critical value $\alpha_{\rm crit}$ slightly shifts to the larger value with decreasing the inflow 
flux, that is $\alpha_{\rm crit} = 4.0$, $4.25$ and $4.75$ for the models with $F_{\rm in}/F_{\rm ref}=1$, 
$1/27$ and $1/1000$ respectively. It is important to stress that the transition value $\alpha_{\rm trans}$ 
does not necessarily coincide with the critical value $\alpha_{\rm crit}$ according to the 1D-HD models, 
even though it has a coincidental match with the critical value in the 2D-MHD models, 
($\alpha_{\rm trans}  \simeq \alpha_{\rm crit} \simeq 5 $ see \S~3.3.). The degeneracy of these 
two values would be dissolved when the input energy is varied in the 2D MHD models.

To specify how the energy inflow pattern affects the dynamics, we examine the model with the 
impulsive inflow flux. The injected energy flux is fixed as $F_{\rm in}/F_{\rm ref} = 1/1000$ 
here. The open squares in Figure~\ref{fig15} represent the 1D-HD models with the impulsive 
inflow. Note that the termination time, at which the energy injection is terminated, is fixed as 
$10\tau$ and is determined to provide the same total input energy as that in the 2D-MHD model 
with the impulsive injection pattern (case B). We find that the different inflow pattern provides 
a different $\alpha$-dependence of the Lorentz factor in the range $\alpha \lesssim \alpha_{\rm crit}$. 
In contrast, there is little difference in the $\alpha$-dependence of the Lorentz factor between 
two models when $\alpha \gtrsim \alpha_{\rm crit}$. These indicate that the energy inflow 
continuously injected into the system does not affect the acceleration of the shock wave as long as 
the initial density profile is sufficiently steep. 

The evolution property of the shock is determined by the total amount of energy injected into 
the system in the duration till the termination time in the impulsive inflow cases. By contrast, 
in the continuous inflow cases, the shock is continuously accelerated without being terminated. 
The steady inflow cases thus provide the larger Lorentz factor than that for the impulsive inflow 
cases in the range $\alpha \lesssim \alpha_{\rm crit}$. 

\begin{figure}
\begin{center}
\scalebox{1}{\rotatebox{0}{\includegraphics{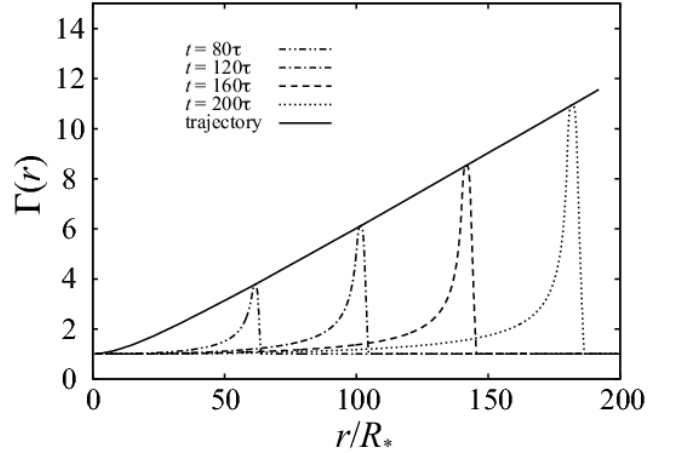}}}
\caption{The time evolution of the Lorentz factor in 1D 
hydrodynamic simulation where $\alpha = 7.25$ and $v_{\rm in}=0.1c$.
A solid line represents the time trajectory of the Lorentz factor 
at the shock surface measured in the laboratory frame.
The meanings of vertical and horizontal axes are same as Fig.9.}
\label{fig16}
\end{center}
\end{figure}

Figure~\ref{fig16}  shows the time evolution of the shock velocity when $F_{\rm in}/F_{\rm ref} = 1$ 
and $\alpha = 7.25$ for the 1D-HD model. The axes have the same meanings as those of 
Figure~\ref{fig9}. The different curves denote the shock profiles at the different phases $t/\tau = 80$, 
$120$, $160$ and $200$ respectively. The solid line represents the trajectory of the Lorentz factor 
at the shock surface. The self-similar property of the relativistic shock in the 2D-MHD model 
is reproduced by the 1D-HD model. The self-similar evolution of the shock observed in the 1D-HD 
models also follows the relation $\Gamma_{\rm sh} \propto r_{\rm sh}$ as is obtained from the 
2D-MHD models. These verify that a purely hydrodynamic process is responsible for the self-similar 
acceleration of the relativistic shock we found in the 2D MHD model. 

\section{Discussion}
\subsection{Comparison with Other Self-Similar Solutions}
\begin{figure}
\begin{center}
\scalebox{1.02}{{\includegraphics{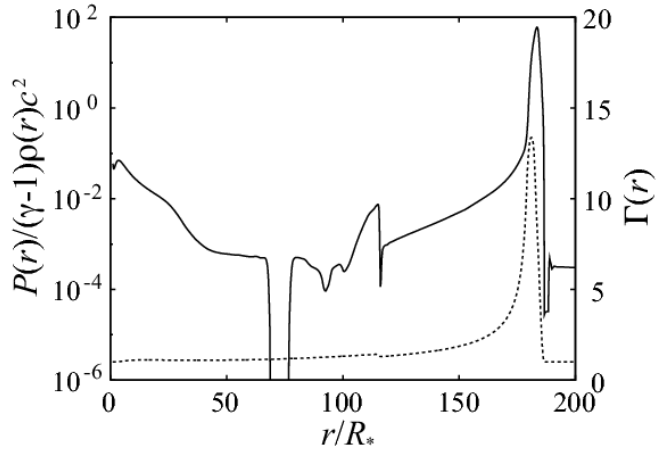}}}
\caption{
The internal energy normalized by the rest mass energy and Lorentz factor of the 
plasma measured when $t=200\tau$ at the equatorial plane in the model where 
$\alpha=7.25$ and $\beta_{0}=3.2$. The left and right vertical axes show the 
internal energy and Lorentz factor respectively.
}
\label{fig17}
\end{center}
\end{figure}
Piran et al. (1993) provides a detailed investigation of the dynamics of a fireball which is 
polluted by radiatively opaque plasma and whose internal energy is significantly greater 
than its rest mass energy. Because of its large opacity and huge internal energy, the extremely 
hot plasma expands adiabatically as a perfect fluid. The homogeneous fireball accelerates 
initially to a ultra-relativistic velocity and then coasts freely after converting all the internal 
energy into the kinetic energy.

The Lorentz factor of the expanding fireball $\Gamma_{\rm fb}$ follows the relation 
$\Gamma_{\rm fb} \propto r_{\rm fb}$ at the early acceleration stage where $r_{\rm fb}$ 
is its radius (Piran et al. 1993). This relation is seemingly the same as the relation describing 
the forward shock obtained from our numerical models. In our model, however, the shock 
propagates not at ultra-relativistic velocity rather at mildly relativistic velocity 
($\Gamma_{\rm sh} \simeq 10$). In addition, unlike the fireball whose internal energy is 
significantly greater than the rest mass energy, the internal energy of the plasma behind the 
shock is less than the rest mass energy in our calculations, as is shown in Figure~\ref{fig17}. 
Here we demonstrate the radial profile of the internal energy and Lorentz factor of the plasma 
measured at the equatorial plane when $t=200\tau$ for the model where $\alpha=7.25$ and 
$\beta_{0}=3.2$. The left vertical axis denotes the internal energy normalized by the rest mass 
energy and the left axis is the Lorentz factor. The physics which characterizes the self-similar 
acceleration of the relativistic shock we found are thus essentially different from that of the 
fireball model.

A spherical blast wave which expands supersonically into the surrounding medium is described 
well by a set of self-similar solutions. The Sedov-Taylor solution is the most famous of these, 
but only describes the case where the velocity is in a non-relativistic regime (Sedov 1959). 
When the released energy is much larger than the total rest mass energy of the explosion products 
and ambient gas, the velocity of the shock becomes relativistic. This type of shock also evolves 
self-similarly according to the Blandford-McKee solution at the ultra-relativistic limit (Blandford \& McKee 1976).

When we consider the shock driven by impulsive energy injection in a gas with the density 
profile $\rho \propto r^{-\alpha}$, it is decelerated in the case $\alpha < 3$ and accelerated 
for $\alpha > 3$ (Shapiro 1979; Waxman \& Shvarts 1993). The evolution of the shock 
follows the second type self-similar solution when the density of the surrounding gas declines 
steeply with radius (Waxman \& Shvarts 1993; Best \& Sari 2000). In the ultra-relativistic case, 
the Lorentz factor of the shock $\Gamma_{\rm sh}$ provides the relation 
$\Gamma_{\rm sh} \propto t^{-m/2}$ with $m = (3-2\sqrt{3})\alpha - 4(5-3\sqrt{3})$, where 
$t$ is the time after the shock initiation (Best \& Sari 2000). The index $-m/2 \simeq 0.77$, $1.0$ 
and $1.23$ when the power law index $\alpha$ is $5$, $6$ and $7$ respectively.

The physical conditions required for the solutions presented above are similar to ours.
However, these are not accountable for the outflow property we found in our study. 
This might be because our self-similar solution is more suited for describing the mildly 
relativistic shock which can not be treated by either non-relativistic or ultra-relativistic 
solutions. We analyze our self-similar solution in more detail by comparing the other 
solutions in a separate paper .

\subsection{Application of Two-Component Outflow Model}
As an application of our two-component outflow model to the realistic astrophysical 
system, we consider the magnetic explosion event in the magnetar system. Giant flares 
from soft gamma-ray repeaters are believed to be powered by the explosive release of 
the magnetic energy on the isolated magnetar (Thompson \& Duncan 1995). It should 
be stressed that the giant flare on the magnetar is followed with an afterglow caused by 
the associated mass ejection event (Cameron et al. 2005; Gaensler et al.2005; Taylor et al. 2005).

Granot et al. (2006) diagnoses the observational properties of radio emitting ejecta 
observed in association with the giant flare on 2004 December 27 from SGR1806-20 
(c.f., Taylor et al. 2005).  It is suggested from their analysis that there is a possibility 
of the ejected matter containing two-types of components of sub-relativistic and 
mildly relativistic velocity. The origin and physics of the mass eruption event has 
not been determined although they give many clues to the model of the magnetar giant 
flare itself. Statistical analysis of giant flaring activity is needed for a better understanding 
of the magnetar.

Although the numerical setting adopted in our model does not correctly capture the 
realistic features of the plasma surrounding the magnetar system, our model creates 
an outflow with a mildly relativistic shock driven component and the subsequent
sub-relativistic magnetically driven component loaded with dense plasma as a natural 
outcome of the magnetic explosion on a compact star.

On the basis of our numerical model, the observed giant flare, energized from the 
magnetic explosion on the magnetar, also powers the two-component outflow as is 
suggested by Granot et al. (2006). We aim to simulate the system of the compact 
star in the numerical setting more suited for the realistic magnetar system as part of our 
future work.
\section{Summary}
We investigate the nonlinear dynamics of the outflow powered by a magnetic 
explosion on a compact star using axisymmetric special relativistic MHD 
simulations. As an initial setting, we set a compact star embedded in the hydrostatic 
gaseous plasma which has a density $\rho (r) \propto r^{-\alpha}$ and is threaded 
by a dipole magnetic field. A longitudinal shearing motion is assumed around the 
equatorial surface, generating the azimuthal component of the magnetic field. 
Subsequent evolution was simulated numerically. Our main findings are summarized 
as follows.

1. Magnetic explosion on the surface of a compact star powers two-component 
outflow expanding into the static medium, i.e, magnetically driven and shock driven 
outflow. At the early evolutionary stage, the magnetic pressure due to the azimuthal 
component of the magnetic field initiates and drives a highly magnetized dense plasma 
outflow. Then the magnetically driven outflow excites a strong forward shock as the 
second outflow component.

2. The expanding velocity of the magnetically driven outflow depends on the Alfv\'en 
velocity at equatorial plane on the surface of the compact star $v_{\rm A,0}$ and 
follows a simple scaling relation $v_{\rm mag} \propto {v_{\rm A,0}}^{1/2}$. 
This scaling relation is accounted for in terms of coupling two balanced equations 
which are reduced from the induction equation and the energy equation.

3. When the initial density profile declines steeply with radius, the outflow-driven 
shock is accelerated self-similarly to relativistic velocity ahead of the magnetically 
driven outflow. The self-similar relation is $\Gamma_{\rm sh} \propto r_{\rm sh}$, 
where $\Gamma_{\rm sh}$ is the Lorentz factor of the fluid measured at the shock 
surface $r_{\rm sh}$. The pure hydrodynamic process is responsible for the 
acceleration of the shock driven outflow and for the self-similarity of the shocked wave. 
\acknowledgments
We thank Hiroyuki R. Takahashi, Tomoyuki Hanawa, Ryoji Matsumoto, 
Takahiro Kudoh, Hiroaki Isobe, Shigehiro Nagataki, and Takaaki Yokoyama for 
their help and useful discussions. We thank Andrew Hillier for his careful reading 
of the manuscript. This work was supported by the Grant-in-Aid for the global 
COE program "The Next Generation of Physics, Spun from Universality and 
Emergence" from the Ministry of Education, Culture, Sports, Science and 
Technology (MEXT) of Japan. This work was partially supported by the 
Short-term Visiting Scholar Program from National Astronomical Observatory 
of Japan. Jin Matsumoto acknowledges support by the Research Fellowship of 
the Japan Society for the Promotion of Science (JSPS).
\appendix
\section{Derivation of the Energy Conservation Law along the Streamline}
When the non-relativistic outflow is mainly polluted by the toroidal magnetic field, the radial outflow velocity should be controlled 
by the following equation of motion, in the spherical coordinate, 
\begin{eqnarray}
\rho \biggl ( \frac{\partial v_{r}}{\partial t} + v_{r} \frac{\partial v_{r}}{\partial r} \biggr ) = - \frac{\partial}{\partial r} \biggl (P + \frac{{B_{\phi}}^2}{8 \pi} \biggr ) - \frac{{B_{\phi}}^2}{4\pi r} - G\frac{\rho M_{*}}{r^2} \;. 
\end{eqnarray}
where all symbols have their usual meanings and the final term of this equation represents a gravitational force of 
the central object with mass $M_{*}$. 
Assuming the steady state, the radial velocity of the magnetically driven outflow should satisfy the steady state 
relation, which is reduced from equation~(A1),
\begin{eqnarray}
\frac{\partial}{\partial r} \biggl ( \frac{1}{2} {v_{r}}^2 \biggr ) = -\frac{1}{\rho} \Biggl [ \frac{\partial}{\partial r} \biggl ( \frac{{B_{\phi}}^2}{8 \pi} \biggr ) + \frac{{B_{\phi}}^2}{4\pi r} \Biggr ] \;, 
\end{eqnarray}
By integrating this equation along the radial streamline from the equatorial surface $R_{*}$ to the arbitrary radius $r$ and dropping the effect of the 
magnetic tension force on the dynamics, we can obtain the radial velocity $v_r (r)$ at the arbitrary radius $r$ as follows; 
\begin{eqnarray}
\frac{1}{2} {v_{r}(r)}^2 = - \int^{r}_{R_{*}} \frac{1}{\rho} \frac{\partial}{\partial r} \biggl ( \frac{{B_{\phi}}^2}{8 \pi} \biggr ) \;{\rm d}r \;. 
\end{eqnarray}
When deriving this equation, it is assumed that the radial velocity at the stellar surface $v_r(r=R_*)$ is negligible small ($< 0.01c$) in 
comparison with the outflow velocity $v_r(r)$ on the basis of our numerical results (see Figure~1), that is $v_r(R_*) \ll v_r(r)$.

\end{document}